\RequirePackage{fix-cm}
\documentclass[fleqn,twocolumn]{svjour3}          
\smartqed  
\usepackage{fix-cm}
\usepackage{graphicx}
\usepackage{float}
\usepackage[caption=false,font=footnotesize]{subfig}
\usepackage{cite}
\usepackage{amsmath}
\usepackage[latin9]{inputenc}
\usepackage{flushend}
\usepackage{balance}
\usepackage{epsfig}

\usepackage{multirow}
\usepackage{array}
\newcolumntype{L}[1]{>{\raggedright\let\newline\\\arraybackslash\hspace{0pt}}m{#1}}
\newcolumntype{C}[1]{>{\centering\let\newline\\\arraybackslash\hspace{0pt}}m{#1}}
\newcolumntype{R}[1]{>{\raggedleft\let\newline\\\arraybackslash\hspace{0pt}}m{#1}}

\usepackage[ruled,vlined,linesnumbered]{algorithm2e}

\DeclareGraphicsExtensions{.eps,.pdf,.png,.ps}
\begin{document}
	
	
	\title{Variability aware Golden Reference Free methodology for Hardware Trojan Detection Using Robust Delay Analysis}
	
	
	
	
\author{Ramakrishna Vaikuntapu \and
		Vineet Sahula\and 
		Lava Bhargava 
}
\vspace{-6ex}
	
\institute{Ramakrishna Vaikuntapu \at
		Department of Electronics \& Telecommunication Engineering \\ 
		Government Engineering College, Bilaspur(C.G) INDIA \\
		\email{vramakrishna409@gmail.com}%
\and
		Vineet Sahula \at
		Department of Electronics and Communication Engineering \\ Malaviya National Institute of Technology Jaipur INDIA \\
		\email{vsahula.ece@mnit.ac.in}%
\and
		Lava Bhargava \at
		Department of Electronics and Communication Engineering \\ Malaviya National Institute of Technology Jaipur INDIA \\
		\email{lavab@mnit.ac.in}              
}
	
\date{Received: date / Accepted: date}
	
\maketitle
	
\begin{abstract}
		
		Many fabless semiconductor companies outsource their designs to third-party fabrication houses. As trustworthiness of chain after outsourcing including fabrication houses is not established, any adversary in between, with malicious intent may tamper the design by inserting Hardware Trojans (HTs). Detection of such HTs is of utmost importance to assure the trust and integrity of the chips. However, the efficiency of detection techniques based on side-channel analysis is largely affected by process variations. In this paper, a methodology for detecting HTs by analyzing the delays of topologically symmetric paths is proposed. The proposed technique, rather than depending on golden ICs as a reference for HT detection,  employs the concept of self-referencing. In this work, the fact that delays of topologically symmetric paths in an IC will be affected similarly by process variations is exploited. A procedure to chose topologically symmetric paths that are minimally affected by process variations is presented. Further, a technique is proposed to create topologically symmetric paths by inserting extra logic gates if such paths do not exist in the design intrinsically. Simulations performed on ISCAS-85 benchmarks establish that the proposed method is able to achieve a true positive rate of 100\% with a false positive rate less than 3\%. In our experimentation, We have considered the maximum of 15\% intra-die and 20\% inter-die variations in threshold voltage ($V_{th}$).
		
\end{abstract}
	
\section{Introduction}
	
\label{sec:intro}
	
	Area, power, and performance have been major concerns for integrated circuit (IC) designers traditionally, whereas the security aspects of ICs have now gained a lot of significance. In the present scenario, the semiconductor industry has adopted a horizontal business model to reduce time-to-market and stay competitive. In such a production model, the IC manufacturing process is spread over the globe and involves many third-party design and fabrication houses. The trustworthiness of such design and fabrication houses is not established. Thus, the security of these chips has become a major concern for system designers and gained the interest of many researchers in recent years. Hardware Trojan (HT) is one such threat to the security and trust of ICs \cite{Spectrum2008-The_Kill_Switch}. Hardware Trojans can be introduced by an adversary at a design house or an untrusted foundry. Depending on the interests of the adversary the HT can cause changes in functionality, denial-of-service, reliability reduction or leaking of secret information \cite{Tehranipoor2010-HT-taxonomy}. Skorobogatov \textit{et al.} have reported an incident of HT in real chips which is a backdoor in military-grade FPGAs \cite{Skorobogatov2012-backdoor-in-military-chip}. Hardware Trojan (HT) can be defined as the unauthorized modification of the original design by an adversary with malicious intent. Such Trojans can be inserted at any abstraction level of IC design and manufacturing process. The detailed taxonomy of hardware Trojans at various abstraction levels can be found in \cite{Tehranipoor2010-HT-taxonomy}. In general, Trojans are intelligently designed to be stealthy, so they do not change the normal functionality of the IC unless they are activated. Such Trojans may not be detected by using conventional functional and structural testing techniques, which are used to test the functionality and detect faults like stuck-at-faults etc. Different methods have been proposed for detecting Trojans inserted at different abstraction levels\cite{Li2015-HTD_survey}. Such detection methods can be classified as design verification phase (pre-silicon), test phase (post-silicon) and run-time \cite{Bao2015-HTD-Temp_Tracking} (post-deployment) methods, based on their applicability at different phases of IC production. The pre-silicon detection methods mainly focus on the Trojans that are inserted by modifying RTL code and/or netlist of the design. These methods use the concepts like unused circuit identification (UCI) \cite{Hicks2010-untrusted-computing-base-UCI}, identification of signals with very low activation probabilities, etc \cite{Zhang2015-Veritrust}.
	
	The test phase i.e post-silicon detection methods aim to detect hardware Trojans in fabricated ICs inserted at untrusted and malicious fabrication foundries. Even though Logic testing \cite{Voyiatzis2016-HT-Triggering-logic, Lesperance2015-HTD-Logic-Test} based approaches for HT detection have been proposed, their efficiency is limited by the stealthy nature of the hardware Trojans and the generation of tests and test execution time. The majority of existing post-silicon methods are devised based on side-channel analysis i.e analyzing the side channel parameters like power, path delay, and electromagnetic measurements, etc. Many existing side-channel analysis (SCA) based HT detection techniques depend on the availability of genuine ICs (golden ICs i.e ICs without any HTs), which are used for generating a reference signature. This signature is then compared against the side-channel signature of suspect IC to decide whether any Trojan exists in the IC under consideration. Obtaining such golden ICs for all the cases is not always feasible, as it involves expensive invasive techniques such as reverse engineering \cite{Bao2015b-reverse_engg_HT_Detection}. Another challenge for side-channel analysis-based methods is the process variation (PV) which impacts the efficiency of detection even if golden ICs are available. To overcome these issues, some golden IC free detection methods have been proposed \cite{Liu2014d-HTD-Golden-free-Statistical, Zheng2015-Semia, Chen2016-HTD-Framework-Statistical-ML, Zhang2018-Golden-free-Processor-HT}, which are based on the concepts of self-referencing and side-channel signature prediction. If the size of inserted Trojan is very small compared to the size of the design, its effect on side-channel parameters may not be differentiable due to process variations (PVs). In the presence of process variation, even if golden ICs are available, direct comparison of the side-channel signatures of the suspect chip with the golden chips may not accurately detect HTs. To mitigate the effects of process variation on side-channel parameters, researchers have proposed a technique known as self-referencing \cite{Self-referencing-HTD}. In this method, the side channel parameter of an IC is compared with the side channel parameter of the IC itself, instead of comparing with the golden signature. The major advantages of self-referencing-based methods are (i) it eliminates the requirement of golden IC and (ii) the effects of inter-die process variation are mitigated.
	
	In the present work, a golden IC-free methodology is proposed to detect Hardware Trojans inserted during fabrication. We have chosen path delay as a medium for Trojan detection because of its specific advantages such as hardware Trojan effect is local to the path and the Trojan affects path delay irrespective of its activation status eliminating the need to activate Trojans. Every IC contains many paths out of which some paths are topologically similar, which we term as symmetric paths. Any two paths which have an equal number of gates of the same type could be considered symmetric paths (see Fig.\ref{fig:topologically_sym_paths}). We analyze the design netlist to identify vulnerable nets which could be potential locations for hardware Trojan insertion. We select a path through such a vulnerable net and we term it as the suspect path. A symmetric path that is topologically similar to the selected suspect path is identified and it is termed as reference path. A symmetric path pair consisting of a suspect and its corresponding reference path is formed. 
	As both the paths in a symmetric path pair pass through the gates of the same type, their delays experience a strong correlation under the inter-die (global or die-to-die) process variation effects. A  symmetric path pair has been selected for each vulnerable net in the design netlist. When the delays of paths in such symmetric path pairs are analyzed by plotting on a two-dimensional space they follow straight lines under inter-die variation (see Fig. \ref{fig:st_line}). If a hardware Trojan has presumably been inserted at a vulnerable net, then the delay of the suspect path will be increased by an amount of HT-induced delay. This in turn deviates the point of path delays away from the expected straight line as shown in Fig.  \ref{fig:st_line}. The distance between the expected line and the point of delays,  due to Trojan-induced deviation has been used to calculate a detection metric (DM) for a selected pair of IC under test. In the presence of intra-die (local or within-die) process variations, the delays of paths in selected symmetric path pairs may tend to deviate from the expected straight line.  Therefore, this detection metric is compared with a pre-defined threshold to separate Trojan inserted ICs from Trojan free ICs. This work mainly focuses on detecting the logical Trojans having payload gate(s). The detection of other types of Trojans like parametric Trojans is out of the scope of this paper.
	
	\subsection{Our Contribution}  
	
	Following are the contributions of our paper.
	
\begin{itemize}
\item We propose a novel path selection algorithm to select symmetric path pairs in a way to mitigate the effects of inter-die variation as well as to minimize the effects of intra-die process variations on hardware Trojan detection accuracy.
\item We present a procedure to create symmetric path pairs if they do not exist in the design intrinsically to cover all vulnerable nodes of the design for efficient Trojan detection.
\item The proposed Trojan detection methodology employs the concept of \textit{self-referencing} wherein the delays of topologically symmetric paths are analyzed for detecting hardware Trojans. The methodology doesn't need a golden reference IC.
\end{itemize}

	The rest of this paper is organized as follows. A brief overview of existing hardware Trojan detection methods is presented in Section \ref{sec:related_work}. Basics of hardware Trojan structure, process variation model, and problem formulation are presented in Section \ref{sec:prelims_prblm_fromltn}. In Section \ref{sec:methodology}, we present the details of the proposed methodology; and  path selection procedure is explained in Section \ref{sec:path_selection_procedure}. Simulation results are reported in Section \ref{sec:simulation_results} and we conclude in Section \ref{sec:conclusion}.

	\section{Related work}
	\label{sec:related_work}
	
	The detailed taxonomy and classification of hardware Trojans have been presented in \cite{Tehranipoor2010-HT-taxonomy}. Several methods based on logic testing have been proposed for detecting such Trojans, but these methods may not always detect active and parametric HTs. The efficiency of these methods is limited due to the large trigger space of hardware Trojan, which is not known to the designer. Thus, side-channel analysis (SCA) based techniques have been considered to be more efficient for detecting hardware Trojans (HTs). Agarwal \textit{et al.} proposed to use side-channel parameter (power) for Trojan detection \cite{Agrawal2007-IC-fingerprinting}. They constructed a fingerprint using golden ICs and compared it with the signatures of suspect ICs for HT detection. Path delays were used to generate the fingerprint of an IC family for HT detection in \cite{Jin2008-path-dly-fingerprint}. They have selected a set of paths that covers the entire design. The path delays of genuine ICs are measured and principle component analysis (PCA) was used to construct the fingerprint. The delays of HT inserted ICs are compared against the fingerprint for HT detection. This method is not scalable for larger designs as it requires a large test time to measure the delays of such a large number of paths covering the entire design. The main limitations of these methods are- (i) the requirement of a set of golden ICs and (ii) the detection accuracy is largely affected by process variations. Authors in \cite{Li2008-HTD_delay_shadow_reg}, a technique was proposed for delay measurement to detect HTs. This technique uses shadow registers to measure the path delays and it can be performed at speed at both test time and run time. This method requires multiple clocks and requires extra registers inserted at the end of each selected path which incurs higher area overhead. The efficiency of delay-based HT detection techniques in the presence of process variations was studied in \cite{Rai2009-delay-based-T-D-Process-variation}. Authors have considered process variation in different transistor parameters like channel length ($L$) and threshold voltage ($V_{th}$). They computed path delays for HT detection by leveraging statistical techniques, even though process variation affects the HT detection accuracy. Authors in \cite{Lamech2012-HTD_delay_variations_REBEL} proposed an embedded test structure by using existing scan structures to measure path delays to detect HTs. In this work, they considered only a small number of arbitrary paths for delay measurement. This technique is not feasible for larger designs as it incurs a large area overhead for on-chip control logic. Authors Xiao et al in \cite{Tehranipoor2013-HTD_Clk_Sweeping} proposed a clock sweeping-based path delay measurement technique for detecting HTs. The authors considered a set of few long paths and used transition delay fault (TDF) test patterns for delay measurements. Statistical techniques were used for signature generation and HT detection. This technique can only be used to detect the HTs inserted in long paths and can not detect the HTs in short paths due to the maximum frequency and power limit of the IC. Authors Cha and Gupta in \cite{Cha2013-Approach-to-select-paths-vectors},  proposed to choose the shortest paths through each possible Trojan location to enhance the effect of HT thus, improving HT detection accuracy in the presence of process variations. By choosing shortest paths only the effect of inter-die variation is minimized and intra-die variation still dominates the detection accuracy. Moreover, this technique cannot perform well when the HT is inserted at such locations through which no shorter path exists. A clock glitching method was proposed to measure path delays and statistical techniques are used to reduce the effects of process variations in \cite{Exurville2015_HTD_resilient_pv}. 
	
	All the aforesaid hardware Trojan detection methods require golden ICs as reference for detecting HTs and they all suffer more or less from both inter-die and intra-die process variations. 
	A self-referencing-based golden-IC free HT detection method using path delays has been proposed in \cite{Norimasa2014-HTD-Symmetry-Breaking-PDly}. It compares the delays of symmetric paths in the design for detecting HTs. This method identifies the symmetries in the design by applying and analyzing the input vector space. This method can not scale up for larger ICs and to do so one has to strike through all possible states which are not always feasible. Authors have suggested developing and use algorithms for identifying symmetries in the design, but they have not reported any such procedure in their work. A pulse propagation technique is proposed by authors in \cite{Deyati2016-HTD-Pulse-propagation} to detect extra capacitance induced by HT on logical paths. Instead of measuring path delays, current sensing circuitry is used for sensing the pulse propagating through the logic paths in the design for HT detection. The efficiency of the method decreases with the increasing number of high fanout nets in the design. A high resolution on-chip embedded test structure called time-to-digital converter (TDC) for measuring path delays and a chip-averaging technique was proposed to detect delay anomalies introduced by HTs in \cite{Ismari2016-HTD-delay-anomalies}. An on-chip self-referencing based HT detection method was proposed in \cite{Zarrinchian2016-latch-based-self-ref-HTD}. Two paths with some pre-defined delay difference are connected to a latch and the output of the latch is considered for HT detection. If a Trojan affects the path having minimum delay then the output of latch changes leading to detection of the HT. This method suffers from the inter-die variation effects as the selected paths may not remain symmetric, resulting in high false-positive rates. Moreover, this method incurs higher area and power overheads for implementing the latches and supporting additional logic such as multiplexers. In \cite{Esirci2017-HTD_correlated_path_delays} an HT detection method was proposed by using the ratio of delays of two paths. To reduce the effects of inter-die variations, shorter paths are selected. The design was simulated for identifying paths in order to minimize the effects of inter-die variations. Simulating the design and identifying such paths is very difficult and incurs longer design times for larger circuits as the number of possible paths increases exponentially. Moreover, this method would not perform well if sufficiently shorter paths do not exist in the design. A variability-aware path selection procedure was reported in \cite{Ramakrishna2018-VLSID2018}. Authors have considered only shorter symmetric paths to mitigate process variation effects, in fact, longer symmetric paths could also perform well in detecting if they are close enough.
	
	Our proposed methodology in this paper attempts to mitigate the effects of inter-die variation as well as tries to minimize the effects of intra-die variation as much as possible enabling more accurate detection of hardware Trojans. However, like any other delay-based detection techniques, the proposed method is also not able to detect Trojans that don't have any impact on path delays.

	\section{Preliminaries and Problem Formulation}
	\label{sec:prelims_prblm_fromltn}
	
	\subsection{Path delay affected by Hardware Trojan}
	\label{subsec:HT_impact_dly}
	
\begin{figure}
		\centering
		\includegraphics[scale=0.5]{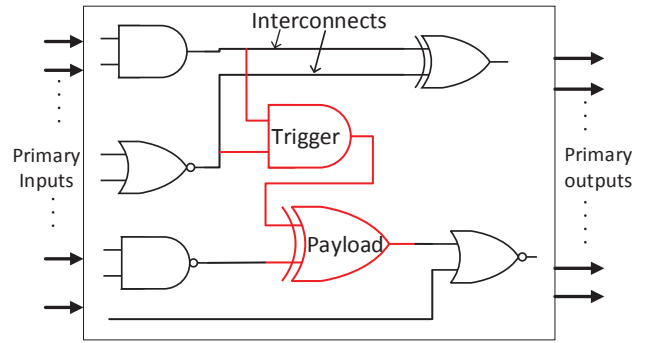}
		\caption{Hardware Trojan structure \cite{Ramakrishna2018-VLSID2018}}
		\label{fig:ht_impact_on_delay}
	\end{figure}

	A Hardware Trojan is constituted by two parts- trigger and payload. The interconnects in the circuit as shown in Fig. \ref{fig:ht_impact_on_delay}, provide inputs to the gates in the HT trigger. However, some additional capacitance,  of this additional fanout, in turn, impacts the delay. The attack is launched by the payload gate, which is inserted by ripping a present net in the original circuit (see Fig. \ref{fig:ht_impact_on_delay}). This net affects the delay of all such paths, which pass through it. The delay is increased by an amount equal to the delay of the payload gate, which usually is higher than the delay caused by the trigger gate. 
	

	\subsection{Hardware Trojan insertion at vulnerable locations}
	\label{subsec:potential_HT_loc}
	
	To defy detection by either testing or side-channel-based techniques, stealthy HTs might be inserted by the attacker. It has been observed that nodes with very low switching activity satisfy the requirements sought after by the attacker \cite{Salmani2009-HTD-improving_DPr_reducing_ActTime, Salmani2016-Analysis-Layout-HT-Insertion, Zou2017-Trigger-HTD}. Hence, we too consider, low switching nets in the circuit as vulnerable locations \cite{Ramakrishna2018-VLSID2018}.
	
	\subsection{Process Variations}
	\label{subsec:PV}
	As the  HTs are very small in size compared to the entire design, their effect on side-channel parameters is not distinguishable from the effects due to process variation. The inter-die variation component affects all transistors of an IC in a similar fashion, whereas, the intra-die variation component is different depending on their physical locations on the IC \cite{Liu2007-PV-spatial-modeling, Chang2003-PV-Spatial-modeling}. Normally, intra-die variations are smaller than inter-die variations. We have considered variations in transistor threshold voltage $V_{th}$ \cite{Ramakrishna2018-VLSID2018}. The variation model considered in this work is shown in (\ref{eq:vth_var})). Further,  the intra-die component $ \Delta v_{intra}(x,y)  $ is divided into- (i) a correlated spatial component and (ii) a random component, as shown in  \eqref{eq:vth_intra}.
	
	\begin{eqnarray}
		\label{eq:vth_var}
		v_{th}(x,y) =  v_{th,nominal} + \Delta v_{inter} + \Delta v_{intra}(x,y) \\
		\label{eq:vth_intra}
		\Delta v_{intra}(x,y) = \Delta v_{spatial}(x,y) + \Delta v_{random}(x,y)
	\end{eqnarray}
	
	Here, $v_{th,nominal}$ is the nominal value of the threshold voltage, and the inter-die variation is $\Delta v_{inter} $. Also, $ \Delta v_{spatial}(x,y) $ is a spatially correlated component, for which a representative model is a correlated multivariate Gaussian random variable; whereas  $\Delta v_{random}(x,y)$ is the random component, modeled as an independent Gaussian random variable.

	\subsection{Topologically symmetric paths} 
	\label{subsec:topologic_sym_paths}
\begin{figure}
		\centering
		\includegraphics[width=0.95\linewidth]{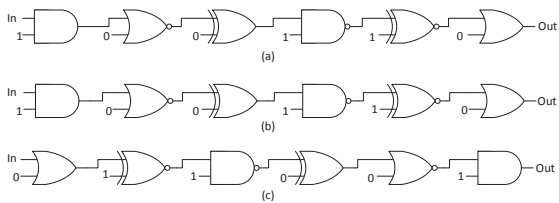} 
		\caption{Symmetric paths. (a) suspect path (b) type--1 reference path (c) type--2 reference path}
		\label{fig:topologically_sym_paths}
	\end{figure}

	In this work, we have considered two paths as topologically symmetric if they pass through similar types of logic gates as shown in Fig. \ref{fig:topologically_sym_paths}. It can be observed that all the three paths shown in Fig. \ref{fig:topologically_sym_paths} are symmetric to each other.   Assuming Fig. \ref{fig:topologically_sym_paths}(a) shows a selected suspect path through a vulnerable net, any one of the two paths shown in Fig. \ref{fig:topologically_sym_paths}(b) and \ref{fig:topologically_sym_paths}(c) can be selected as a reference path for the selected suspect path as these paths are topologically symmetric to path shown in Fig. \ref{fig:topologically_sym_paths}(a). We call the path shown in Fig. \ref{fig:topologically_sym_paths}(b) as type--1 reference path and the path shown in Fig. \ref{fig:topologically_sym_paths}(c) as type--2 reference path. The delay of the paths shown in Fig. \ref{fig:topologically_sym_paths} may or may not be equal as it also depends on the fanout and interconnects delays experienced by gates in the respective paths of the design.

\begin{figure}
\centering
\includegraphics[scale=0.9]{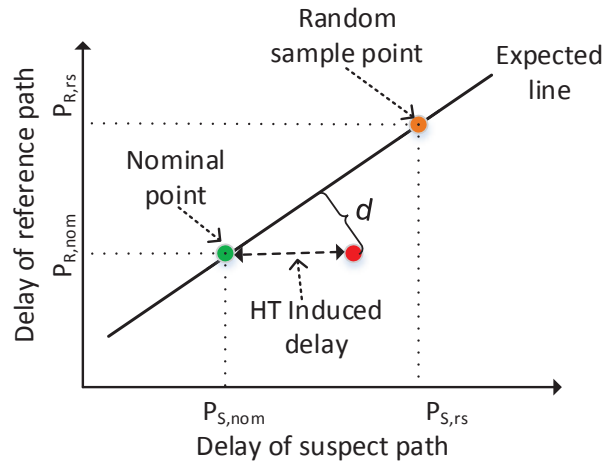}
\caption{Deviation from expected straight line}
\label{fig:st_line}
\end{figure}

	\begin{table*}
		\centering
		\caption{Correlation between path delays of a symmetric path pair for different delay models}
		\label{table:correlation_types_results}
		\begin{tabular}{|C{1.7cm}|c|c|c|c|c|c|}
			\hline
			Symmetric path pair & \multicolumn{3}{c|}{{type--1}}  &  \multicolumn{3}{c|}{{type--2}} \\
			\hline
			Delay model & $rise\_delay$ &  $fall\_delay$ & \large{$\frac{rise\_delay + fall\_delay}{2}$} & $rise\_delay$ & $fall\_delay$ &  \large{$\frac{rise\_delay + fall\_delay}{2}$} \\
			\hline
			Correlation (\%) & 100 & 100 & 100 & 98.712 & 97.687 &  99.997 \\
			\hline
		\end{tabular}
	\end{table*}

	\begin{table*}
		\caption{Correlation between path delays of a symmetric path pair for different path lengths and fanout difference}
		\label{table:correlation_all_results}
		\centering
		\begin{tabular}{|c|c|c|c|c|c|c|c|c|c|} 
			\hline
			\multirow{2}{*}{Symmetric path pair}& \multicolumn{4}{c|}{{path length}}& \multicolumn{5}{c|}{fanout difference (fod)} \tabularnewline \cline{2-10} &9&12&15&18&0&3&6&9&12 \tabularnewline
			\hline
			Correlation (\%)&99.973&99.984&99.995&99.965&100&99.996&99.993&99.974&99.922 \\
			\hline
		\end{tabular}
	\end{table*}

	The fundamental difference between type--1 and type--2 reference paths can be explained by propagating a transition through these paths concerning the suspect path. In the type--1 path the signal transition propagates through gates of the same type in the same order as that of the suspect path. Whereas, in type-2 paths, the signal transition propagates through similar gates but in a different order as that of the suspect path. For example, the XOR gate in Fig. \ref{fig:topologically_sym_paths}(b) experiences the same transition as that of the XOR gate in Fig. \ref{fig:topologically_sym_paths}(a) when an identical transition is applied at the input pin of the two paths. Whereas, the XOR gate in Fig. \ref{fig:topologically_sym_paths}(c) experiences the opposite transition as that of the XOR gate in Fig. \ref{fig:topologically_sym_paths}(a) when an identical transition is applied at the input pin of the two paths. Further, we identify a symmetric path pair as a type--1 (type--2) if the selected reference path is of type--1 (type--2). In our work, the path delay has been considered differently for paths in type--1 and type--2 symmetric path pairs. We assume the delay of a path for propagating a rising (falling) transition at its input port through it is represented by $rise\_delay$ ($fall\_delay$).

	In type--1 symmetric path, every gate in the reference path experiences the same transition (rise or fall) concerning the same type of gate in the suspect path. Thus, the delays of the suspect path and the reference path are highly correlated for the identical transition applied at their input ports. Therefore, for paths in type--1 symmetric path pairs the path delay has been considered as either rise delay or fall delay as shown in (\ref{eq:type-1-delay}). i.e. if $rise\_delay$ ($fall\_delay$) is considered as path delay for suspect path then $rise\_delay$ ($fall\_delay$) of the reference path has to be considered as path delay.

	\begin{equation} 
		\centering
		\label{eq:type-1-delay}
		\text{path\_delay} =\begin{cases} 
			\text{rise\_delay} & \text{or}\\ 
			\text{fall\_delay} 
		\end{cases} 
	\end{equation}
	
	In type--2 symmetric path pairs, some gates of reference path may experience different transitions concerning the gates of the same type in the suspect path. To compensate for the effects of such transition difference on the correlation between the suspect and reference path delays, the average of rise and fall delays has been considered as the path delay. The rationale behind the average delay is when rising and falling transitions are applied at the inputs of both suspect and reference paths, all the gates in these paths also experience both rising and falling transitions. Therefore, for paths in type--2 symmetric path pairs the path delay has been considered as the average of rise and fall delays as shown in \eqref{eq:type-2-path_delay_model}.
	
	\begin{equation}
		\label{eq:type-2-path_delay_model}
		\text{path\_delay} = \frac{\text{rise\_delay + fall\_delay}}{2}
	\end{equation}
	
\begin{figure*}%
\centering
\subfloat[]{%
		\includegraphics[scale=0.4]{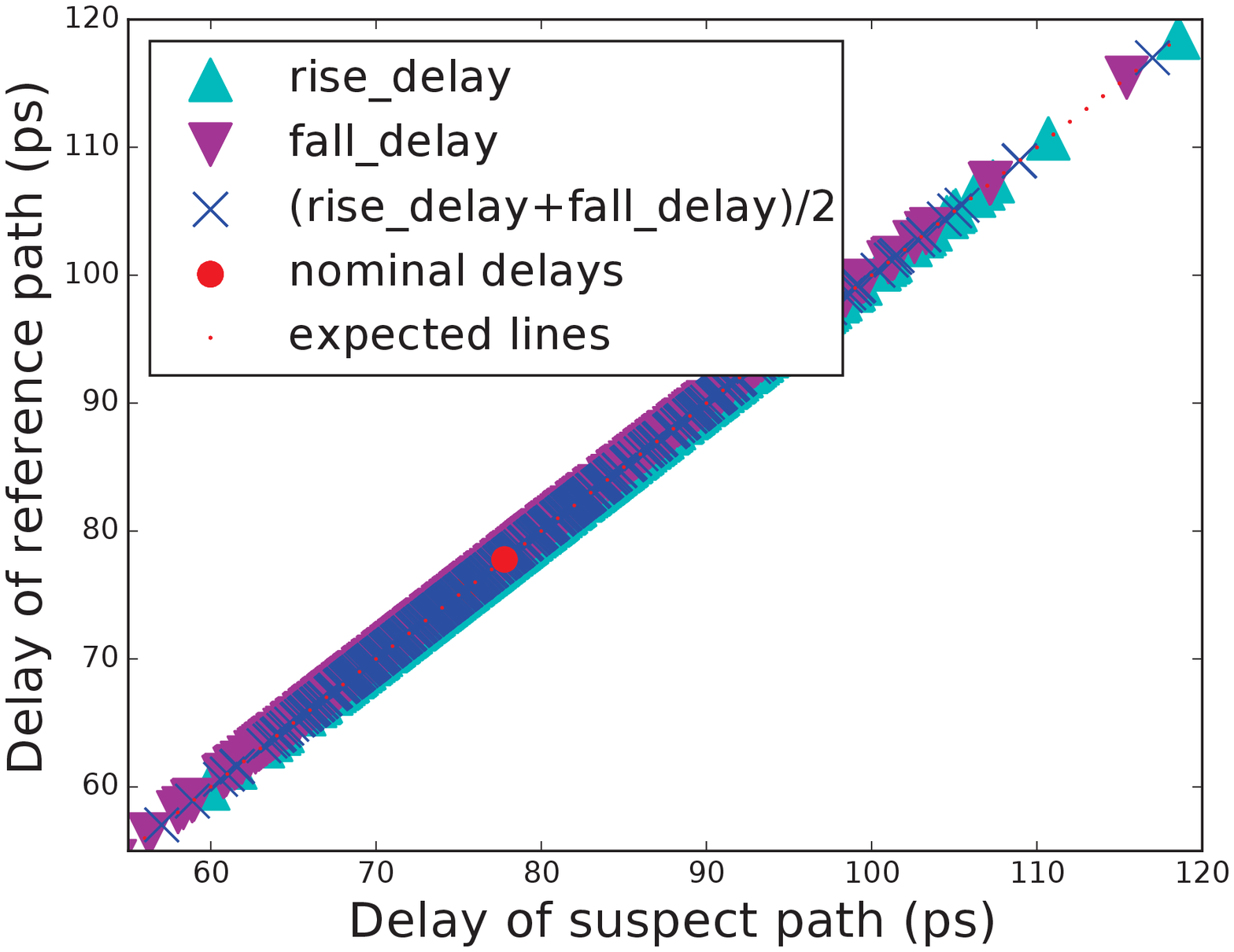}\label{fig:type-1_dly_model}}
\hfill
\subfloat[]{%
		\includegraphics[scale=0.4]{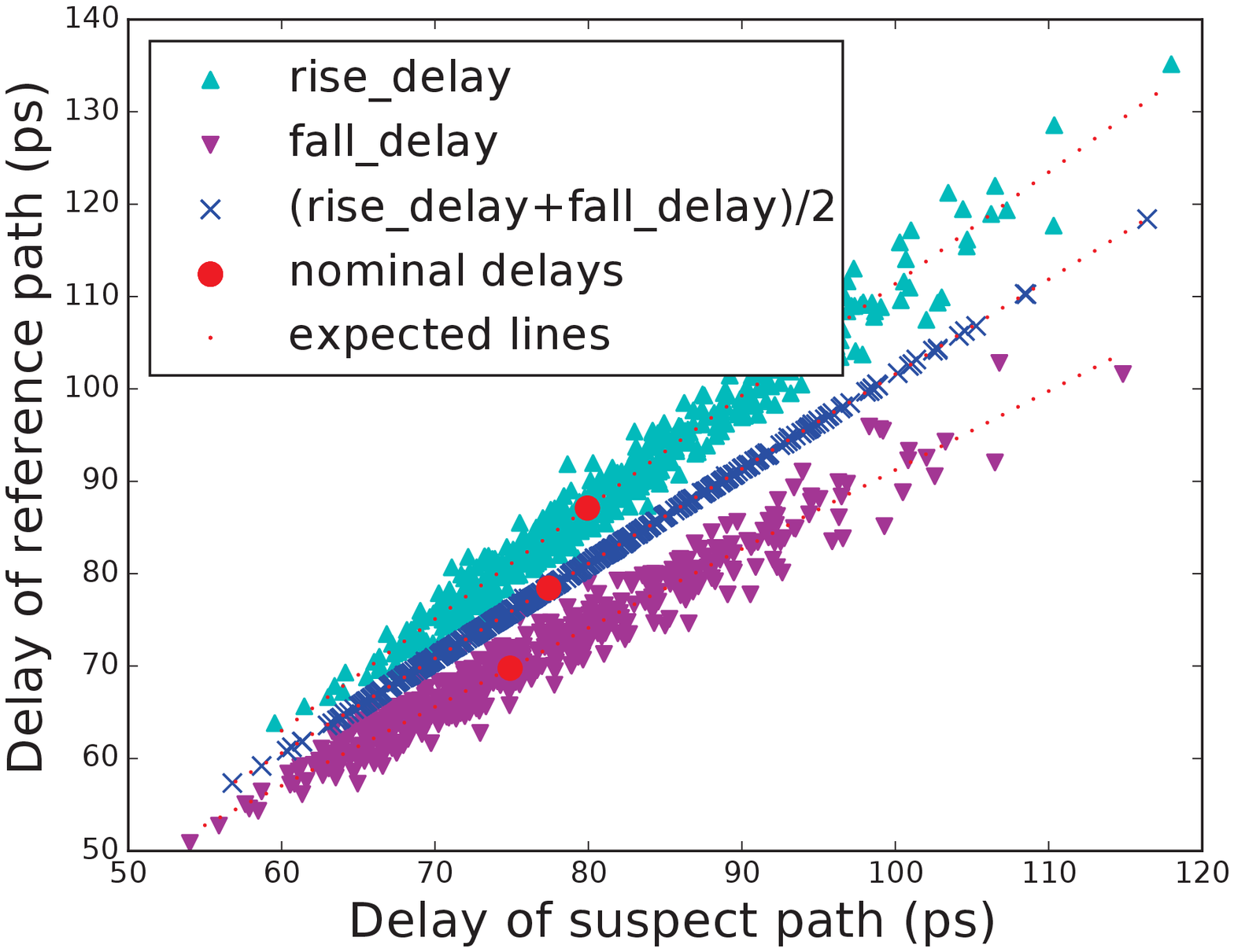}\label{fig:type-2_dly_model}}%
\caption[]{Correlation between suspect path and reference path delays under inter-die variations for different delay models considered
			\subref{fig:type-1_dly_model} for paths in type--1 symmetric path pair
			\subref{fig:type-2_dly_model} for paths in type--2 symmetric path pair
}%
\label{fig:types_sym_path_correlation}
\end{figure*}

	The correlation between different delays of suspect path shown in Fig. \ref{fig:topologically_sym_paths}(a) and reference path shown in Fig. \ref{fig:topologically_sym_paths}(b) of a type--1 symmetric path pair under inter-die variation is shown in Fig. \ref{fig:types_sym_path_correlation}\subref{fig:type-1_dly_model}. It can be observed that the rise, fall and average of rise and fall delays of suspect and reference paths  are highly correlated under inter-die variation effects. Therefore, either $rise\_delay$ or $fall\_delay$ an be considered as the path delay for the paths in type--1 symmetric path pairs. The percentage correlation coefficient between the delays of paths in type--1 symmetric path pair is shown in Table \ref{table:correlation_types_results}.
	
	Similarly, a type--2 symmetric path pair has been formed by considering the path shown in Fig. \ref{fig:topologically_sym_paths}(a) as suspect path and the path shown in Fig. \ref{fig:topologically_sym_paths}(c) as reference path. The correlation between the suspect and reference path delays of the type--2 symmetric path pair for different delay models considered under inter-die variations is shown in Fig. \ref{fig:types_sym_path_correlation}\subref{fig:type-2_dly_model}. The percentage correlation coefficient between the rise, fall, and the average of both delays of the suspect and the reference paths is shown in Table \ref{table:correlation_types_results}. It is observed that the correlation between rise delays of the suspect path and the reference path is 98.712\%. Similarly, the correlation between the fall delays is 97.687\%. Whereas, the correlation between the average rise and fall delays of suspect and reference paths of a type--2 symmetric path pair is 99.997\%. Therefore, to maintain a high correlation, the average of rise and fall delays as shown in (\ref{eq:type-2-path_delay_model}), has been considered as the path delay for the paths in type--2 symmetric path pairs.

\begin{figure*}%
\centering
\subfloat[]{\includegraphics[scale=0.4]{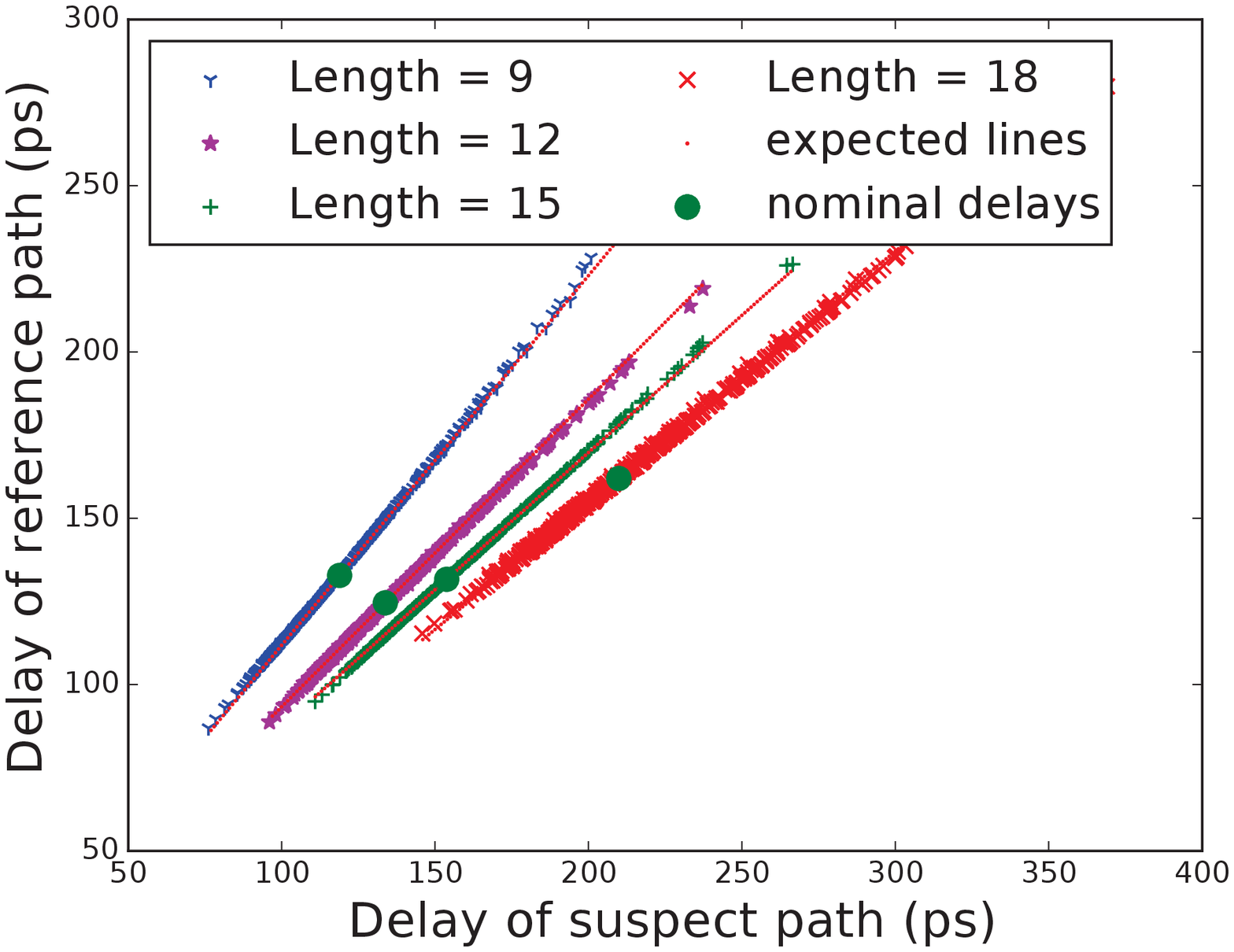}\label{fig:different-path-lengths}}%
\hfill
\subfloat[]{\includegraphics[scale=0.4]{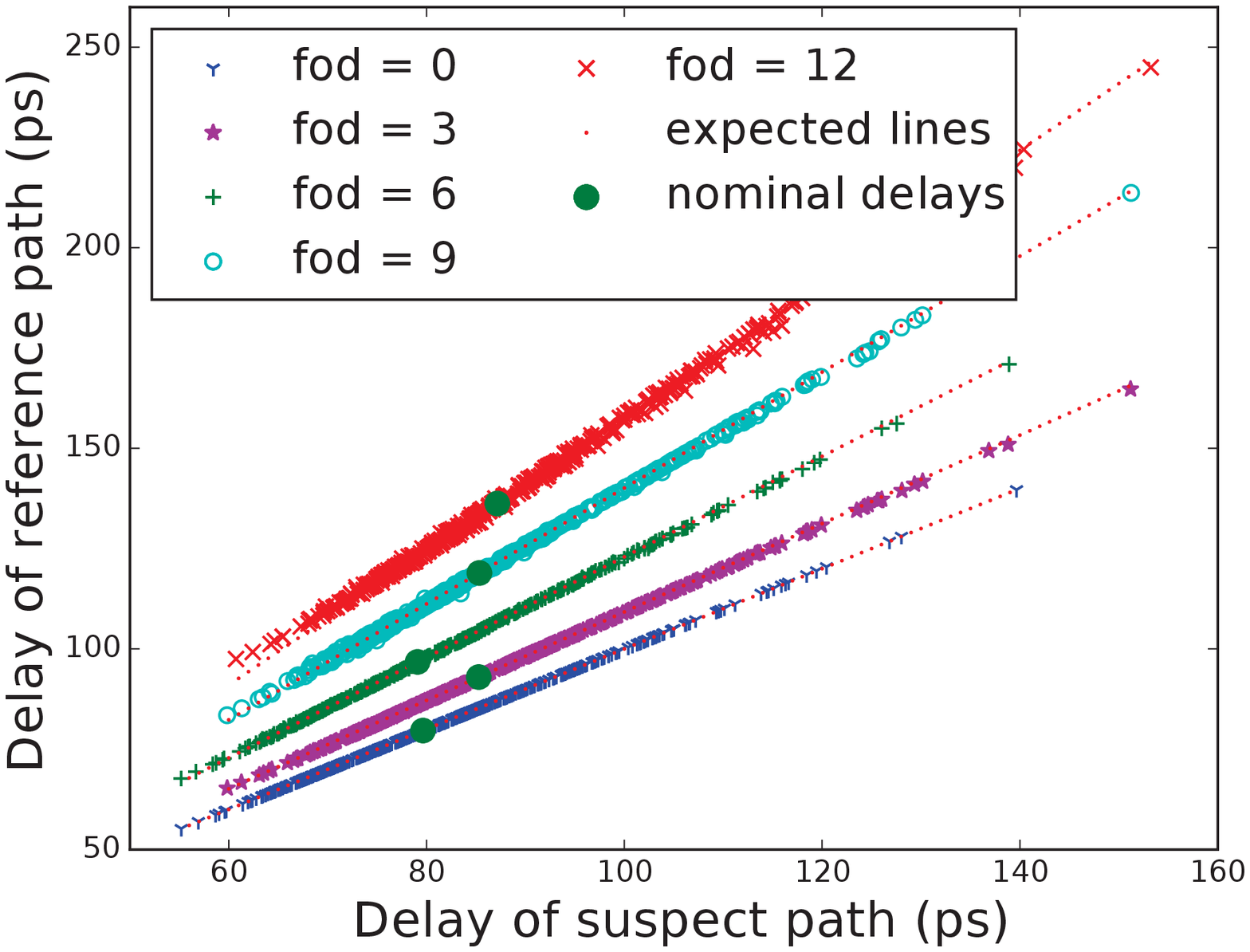}\label{fig:different-fanout-differences}}%
\caption[]{Effect of inter-die process variations on delays of suspect and reference paths along the expected line
	\subref{fig:different-path-lengths} for different path lengths
	\subref{fig:different-fanout-differences} 
for different fanout differences}%
\label{fig:pl-fod-sym_path_correlation}%
\end{figure*}

	The effect of inter-die variation on the delays of paths of symmetric path pairs with different path lengths has been shown in Fig. \ref{fig:pl-fod-sym_path_correlation}\subref{fig:different-path-lengths}. We have considered the number of gates in a path as the path length. The path pairs with lengths 9 and 12 are extracted from c1908 and  c2670 benchmark circuits respectively. The path pairs with path lengths 15 and 18 are selected from the c7552 benchmark circuit. It is observed that the delays are as expected for all the path lengths, thus leading to the inference that the effect of inter-die variation is mitigated irrespective of the length of symmetric paths considered. We can even observe that the nominal delays of suspect and reference paths are not equal even though they pass through the same type of gates. This is due to the difference in fanout of each gate and the length of interconnects between the gates in these paths.

	We have analyzed the effect of fanout difference on the correlation between the delays of the suspect and the reference paths shown in Fig. \ref{fig:topologically_sym_paths}(a) and Fig. \ref{fig:topologically_sym_paths}(b) respectively. The fanout difference (fod) is considered as the difference between the total fanout of suspect and reference paths.  We have connected INVX1\_RVT gate to the nets in the suspect and reference paths randomly to create the fanout difference. For example, to create a fanout difference of 3, we have connected 2 INVX1\_RVT gates to the randomly chosen nets of the suspect path and 5 INVX1\_RVT gates to the nets of reference path randomly. The correlation between path delays of suspect and reference paths under inter-die variation for several fanout differences has been shown in Fig. \ref{fig:pl-fod-sym_path_correlation}\subref{fig:different-fanout-differences}. It can be observed that the fanout difference only affects the slope of the expected lines but does not have any impact on the correlation between path delays under the effects of inter-die variation. Table \ref{table:correlation_all_results} presents the correlation between the delays of symmetric paths under inter-die variation for different path lengths and fanout differences. It can be observed that the percentage correlation coefficient is greater than 99.9\% in all the considered cases. Therefore, we infer that the inter-die process variation effects on hardware Trojan detection can be mitigated by selecting topologically symmetric paths, irrespective of their type, length, and fanout differences.
	
	Let us consider Fig. \ref{fig:st_line}, the nominal delays of suspect path and reference path are represented by $P_{s,nom}$ and $ P_{r,nom}$ respectively i.e. these are the delays at nominal transistor threshold voltage $v_{th,nom}$. $ P_{s,rs}$ and $ P_{r,rs}$ are the delays of suspect and reference paths respectively for a random sample due to inter-die variation i.e. these are the delays at transistor threshold voltage $v_{th,nom}+\Delta v_{inter}$. where $\Delta v_{inter}$ is a random sample from inter-die variation distribution modeled using (\ref{eq:vth_var}), which will be same for all the transistors in the two paths as explained in Section  \ref{subsec:PV}. The expected straight line equation can be determined by using the nominal and random sample delay  points as shown in (\ref{eq:st_line_eq1}).
	
	\begin{equation}
		\label{eq:st_line_eq1}
		y-P_{r,nom} = \frac{P_{r,rs}-P_{r,nom}}{P_{s,rs}-P_{s,nom}}(x-P_{s,nom})
	\end{equation} 
	
	By rearranging the terms in (\ref{eq:st_line_eq1}) we get the equation of the expected straight line as follows, where $\alpha = (P_{r,rs}-P_{r,nom})/(P_{s,rs}-P_{s,nom}) $ and $\beta=(P_{r,nom}-\alpha P_{s,nom})$.
	\begin{equation}
		\label{eq:st_line_eq2}
		\alpha x-y+\beta =0
	\end{equation}

	Assuming a hardware Trojan gate has presumably been inserted in the suspect path and it, in turn, induces some extra delay. The delays of suspect path after  Trojan insertion and reference path are represented by $P_{s}$ and $P_{r}$, respectively. Due to HT-induced extra delay, the point ($P_{s}, P_{r}$) deviates from the expected straight line. This deviation causes the distance $d$ between the point ($P_{s}, P_{r}$) and the expected line. This distance $d$ has been used to calculate the detection metric ($DM$) to detect hardware Trojan. The distance $d$ can be calculated as per (\ref{eq:distanced_d}).
	
	\begin{equation}
		\label{eq:distanced_d}
		d = \frac{1}{\sqrt{1+\alpha^2}}\left(\left|\alpha P_{s}-P_{r}+\beta\right|\right)
	\end{equation}
	
	The detection metric  $DM$ is calculated as the normalized distance \cite{Ramakrishna2018-VLSID2018} from the expected straight line as shown in  (\ref{eq:DM}).
	
	\begin{equation}
		\label{eq:DM}
		DM = \frac{d}{\sqrt{(P_{s,nom})^2 + (P_{r,nom})^2}}
	\end{equation}

	Under the influence of process variations, the delays of the suspect and the reference paths may deviate from their nominal values. These delays follow the expected line under inter-die variation as the paths are topologically symmetric. The intra-die variation effects cause the deviation of the delay point away from the expected line as it affects transistors in the symmetric paths differently depending on their physical location on the die. Thus, ideally, under no process variations, the value of $DM$ should be zero. It can be understood that even in the presence of inter-die variation also the $DM$ is zero as the delay points lie on the expected line. But, under intra-die variation as the delay points deviate from the expected line, the $DM$ becomes non-zero i.e. $DM>0$. Therefore, the computed detection metric $DM$ is compared with a pre-defined threshold ($DT$) to separate Trojan inserted ICs from Trojan free ICs. The detection threshold $DT$ can be obtained by performing Monte-Carlo simulations with reliable process variation models. The hardware Trojan detection problem can be formulated as
	
	\begin{equation} 
		\label{eq:HTD_problem}
		\text{IC\ under\ test} =\begin{cases} 
			\text{is Trojan free}  & \text{if } DM<DT\\ 
			\text{has Trojan inserted} & \text{otherwise}
		\end{cases} 
	\end{equation}

	\begin{figure*}
		\centering{%
			\includegraphics[width=17.5cm]{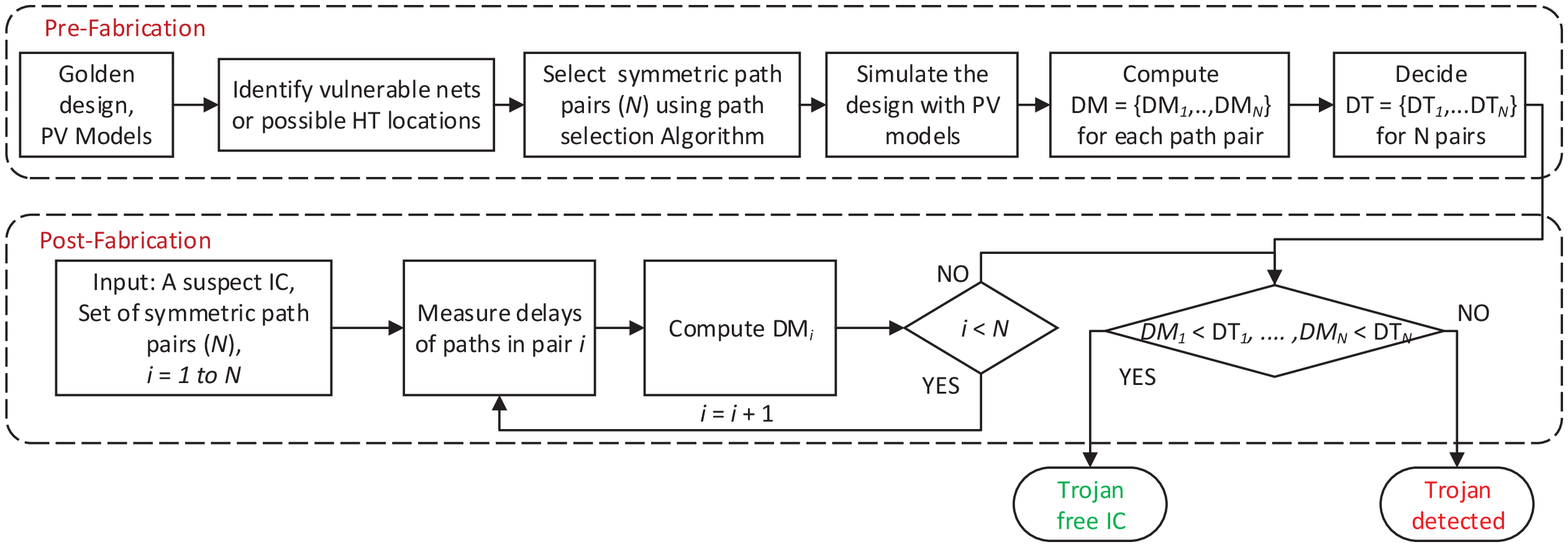}%
		}
		\caption{Proposed Hardware Trojan Detection Methodology}
		\label{fig:methodology}
	\end{figure*}

	\section{Detection Methodology}
	\label{sec:methodology}
	
	In this section, we discuss our proposed methodology in detail. The proposed methodology consists of two stages i.e (i) the Pre-fabrication stage and (ii) the Post-fabrication stage. The complete flow of the proposed HT detection methodology is shown in Fig. \ref{fig:methodology}.
	
	\subsection{Pre-fabrication stage}
	\label{subsec:pre_fab_stage}
	We assume that the golden model of the design is available since we have considered the threat model of HT insertion at untrusted foundry i.e. the HT is assumed to be inserted after design sign-off while fabrication. The design netlist has been analyzed to find vulnerable nets, which are considered to be potential locations of HT insertion. We have considered low activity nets as the vulnerable locations for HT insertion as explained in Section \ref{subsec:potential_HT_loc}. The set of vulnerable nets $\{N_{v}\}$ has been identified using the simulation of the golden netlist of the design over a large number of random test vectors. A symmetric path pair (SPP) consisting of a suspect path and a reference path is selected for each vulnerable net in the design. The path selection algorithm which is used to select suspect and reference paths is explained in Section  \ref{sec:path_selection_procedure}. A set of symmetric path pairs $ \{ N \}$ is formed by combining a suspect path and its corresponding reference path for each net $n$ in $\{N_{v}\}$. The nominal delays of suspect path and reference path of $i^{th}$ symmetric path pair corresponding to $i^{th}$ vulnerable net in $\{N_{v}\}$ are represented by $P_{s,nom}^{i}$ and $P_{r,nom}^{i}$ respectively. The detection metric corresponding to $i^{th}$ pair $DM_{i}$ is calculated as the normalized distance from expected straight line using (\ref{eq:distanced_d}) and (\ref{eq:DM}). Monte-Carlo simulations are performed using reliable process variation models provided by the foundry to generate a set of Trojan-free IC instances. The metric $ DM $ = $\{ DM_{1},DM_{2},.....,DM_{N}\}$ corresponding to all selected symmetric path pairs of the design is calculated for all Trojan free IC instances. Using this data a detection threshold is decided for each symmetric path pair. The set $DT$ = $\{DT_{1},DT_{2},.....,DT_{N}\}$ contains these threshold values. This $DT$ is used in the post-fabrication stage for HT detection.
	
	\subsection{Post-fabrication stage}
	\label{post_fab_stage}
	The post-fabrication stage involves the measurement and analysis of delays of the paths present in selected symmetric path pairs. The path delays of all $2N$ paths in $N$ pairs are measured for each suspected IC. Techniques presented in\cite{Li2008-HTD_delay_shadow_reg,Tehranipoor2013-HTD_Clk_Sweeping} can be used to measure the delays of selected paths.  The $DM$ of each pair is calculated using (\ref{eq:distanced_d}) and (\ref{eq:DM}). Each $DM_{i}$ is compared with pre-defined detection threshold  $DT_{i}$ in order to infer whether the IC under test is Trojan free as shown in (\ref{eq:HTD_methodology}).
	
	\begin{equation}
		\label{eq:HTD_methodology}
		\text{IC\ under\ test} =  
		\begin{cases} 
			\text{is\ Trojan\ free\ } \hspace{1.05cm} if\ DM_{i}<DT_{i}\\ \hspace{3.95cm} \forall\ 1<i\leq N\\
			\text{has\ Trojan\ inserted\ } \   \  \ \ \ \ \    \ otherwise 
		\end{cases}
	\end{equation}
	
	\section{Path selection Procedure}
	\label{sec:path_selection_procedure}
	
	The proposed path selection procedure and reference path creation technique are described in this Section.

	\subsection{Path selection algorithm}
	\label{subsec:path_sel_algo}
	
	The proposed path selection technique is presented in Algorithm \ref{algo:path_sel_algo}. This algorithm takes a set of vulnerable nets ${N_{v}}$ and the golden design netlist as inputs and returns a topologically symmetric path pair (SPP) consisting of a suspect path and its corresponding reference path for each vulnerable net in the design. All sensitizable paths passing through the vulnerable net $n_{i}$ are collected in set $S_{all\_paths}^{i}$. We have used Boolean satisfiability (SAT) based techniques to decide whether a path is sensitizable \cite{MarquesFelipeS2005-SAT-Falepath-Detection, Lu2008-SAT-based-PD-ATPG, Sauer2016-PHAETON-SAT}. For the sake of clarity and simplicity, wherever we refer to a path in the remainder of the paper, it is to be considered as a sensitizable path. All the paths that have symmetric paths (i.e. either type-1 or type-2) are collected in set $S_{suspect}^{i}$. The nets which have at least one symmetric path for at least one path out of all paths passing through them are collected in set $N_{covered}$ and called covered nets. If there are no symmetric paths for at least one path in $S_{all\_paths}^{i}$ then we call such nets as uncovered nets. We need to create a symmetric path to be used as a reference path for each of such nets. The procedure as explained in Algorithm \ref{algo:ref_path_cretn_algo}, is used for reference path creation. All possible suspect paths through the net $n_{i}$ are collected in set $S_{suspect}^{i}$ and all the paths symmetric to the path $P_{suspect,j}^{i}$ in $S_{suspect}^{i}$ are collected in $S_{symmetric,j}^{i}$.
	
\begin{figure}
		\centering
		\includegraphics[scale=0.6]{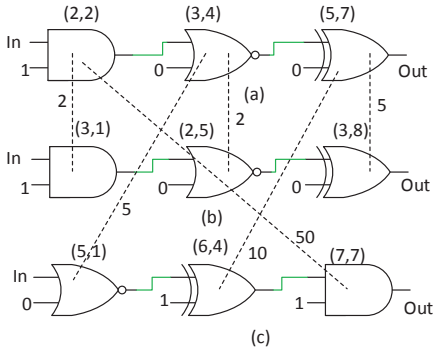}
		\caption{Example for ranking symmetric path pairs according to the locations of gates in the layout (a) suspect path (b) symmetric path with rank = 3 i.e. $(2+2+5)/3$ (c) symmetric path with rank = 21 i.e. $(50+5+10)/3$}
		\label{fig:path_rank}
	\end{figure}

	After finding out possible suspect paths and corresponding symmetric paths for each vulnerable net, the final layout of the netlist is prepared. The physical location information of each gate is collected from the layout of the netlist and is used for selecting the nearest symmetric path pair to exploit the spatial correlation component of intra-die variation. We compute the rank of a symmetric path pair as the average of distances between similar gates in a symmetric path pair. The procedure for computing the rank of a symmetric path pair is presented in Algorithm \ref{algo:path_rank_algo}. The suspect path and corresponding symmetric path are represented by $P_{suspect,j}^{i}$ and $P_{symmetric,k}^{i}$ respectively. The gates in paths $P_{suspect,j}^{i}$ and $P_{symmetric,k}^{i}$ are collected in sets $G_{susp}$ and $G_{sym}$, respectively. Assume $g_{l}$ and $g_{m}$ are similar type gates present in $G_{susp}$ and $G_{sym}$ and their locations  on the layout of the design are represented by ($x_{l}$, $y_{l}$) and ($x_{m}$, $y_{m}$), respectively.
	The Euclidean distance $dist_{q}$ between these two gates $g_{l}$ and $g_{m}$ is calculated as shown in line 8 of Algorithm \ref{algo:path_rank_algo}. The rank of the symmetric path pair is computed by taking the average of the distances $dist_{q}$ for all the gates in $G_{susp}$ as shown in line 10 of algorithm \ref{algo:path_rank_algo}. Here, $|G_{susp}|$ represents the cardinality of the set $G_{susp}$ which is nothing but the number of gates in the suspect path. An example illustrating the Algorithm \ref{algo:path_rank_algo} is presented in Fig. \ref{fig:path_rank}. A pair that has nearest symmetric paths i.e with lowest rank is selected and a symmetric path pair $SymmetricPathPair_{i}$ corresponding to net $n_{i}$ is formed as $ (p_{susp}^{i}, p_{ref}^{i})$.

	\begin{algorithm}
		\caption{Path selection algorithm}
		\label{algo:path_sel_algo}
		\DontPrintSemicolon
		\SetAlgoLined
		\SetKwInOut{Input}{Input}
		\SetKwInOut{Output}{Output}
		\Input{ Netlist, $N_{v}$ = set of vulnerable nets}
		\Output{Symmetric path pairs}
		\BlankLine
		
		\ForEach{net $n_{i}$ in $N_{v}$}{    
			
			\BlankLine
			$S_{all\_paths}^{i}$ = set of all sensitizable paths passing through the net $n_{i}$\;
			$S_{suspect}^{i}$ = $S_{all\_paths}^{i}$\;
			\ForEach{path $P_{suspect,j}^{i}$ in $S_{all\_paths}^{i}$}{
				$S_{symmetric,j}^{i}$ = set of paths that are symmetric (type-1 or type-2) to $P_{suspect,j}^{i}$ and not passing through net $n_{i}$\;
				\If{$|S_{symmetric,j}^{i}| = 0 $}{
					{\scriptsize{ \tcc{If there are no symmetric paths for path $P_{suspect,j}^{i}$ then drop it from the suspect path list $S_{suspect}^{i}$} } }
					$S_{suspect}^{i}$ = $S_{suspect}^{i}$ - $P_{suspect,j}^{i}$\;
				}
			}
			
			\uIf{$|S_{suspect}^{i}| = 0 $}{
				
				{\scriptsize { \tcc{If there are no symmetric paths for at least one path in $S_{all\_paths}^{i}$ then create a symmetric path } } }
				\textit{refpathgen}($n_{i}, S_{all\_paths}^{i}$) 
			}\Else{Add net $n_{i}$ to the set of covered nets $N_{covered}$} 
		}
		Perform physical synthesis of (modified) netlist for generating layout of the design \;
		Prepare the physical location information of each gate in the netlist\;
		\ForEach{net $n_{i}$ in $N_{covered}$}{ 
			$rank\_initial = \infty $\;
			\ForEach{path $P_{suspect,j}^{i}$ in $S_{suspect}^{i}$}{
				\ForEach{path $P_{symmetric,k}^{i}$ in $S_{symmetric,j}^{i}$}{
					Compute the rank of the path pair $P_{suspect,j}^{i}$ and $P_{symmetric,k}^{i}$ which is dependent on the distance between the two paths.\;
					$rank_{jk}$ = \textit{rank}($P_{suspect,j}^{i},P_{symmetric,k}^{i}$)\;
					\If{$rank_{jk} < rank\_initial$ }{ {\scriptsize{\tcc{selects the path pair with lowest rank i.e the nearest symmetric path pair}} }
						$P_{susp}^{i}=P_{suspect,j}^{i}$ and
						$P_{ref}^{i}=P_{symmetric,k}^{i}$\;
						$rank\_initial=rank_{jk}$\;
					}
				}
			}
			$SymmetricPathPair_{i} \leftarrow ( P_{susp}^{i}, P_{ref}^{i} )$\;   
		}
	\end{algorithm}

	\begin{algorithm}
		\caption{Calculation of rank of a path pair -- \textit{rank}()} 
		\label{algo:path_rank_algo}
		\DontPrintSemicolon
		\SetAlgoLined
		\SetKwInOut{Input}{Input}
		\SetKwInOut{Output}{Output}
		\Input{A pair of paths $P_{suspect,j}^{i}$, $P_{symmetric,k}^{i}$  }
		\Output{Rank of the path pair}
		$G_{susp}$ = Set of all gates in path $P_{suspect,j}^{i}$\;
		$G_{sym}$ = Set of all gates in path $P_{symmetric,k}^{i}$\;
		\ForEach{ gate $g_{l}$ in $G_{susp}$ }{
			\ForEach{ gate $g_{m}$ in $G_{sym}$ }{
				\If{$g_{l}$=$g_{m}$}{{\scriptsize { \tcc{If both gates are of same type} } }
					location of $g_{l}$ on layout is ($x_{l},y_{l}$)\;
					location of $g_{m}$ on layout is ($x_{m},y_{m}$)\;
					$dist_{q} = \sqrt{(x_{l}-x_{m})^2+(y_{l}-y_{m})^2}$
				}
				$rank = \frac{\sum_{q=1}^{|G_{susp}|}dist_{q}}{|G_{susp}|} $\;
			}
		}
		return $rank$\;
	\end{algorithm}

	\subsection{Reference path creation}
	\label{subsec:Ref_path_creation}
	
	\begin{figure}
		\centering
		\includegraphics[scale=0.6]{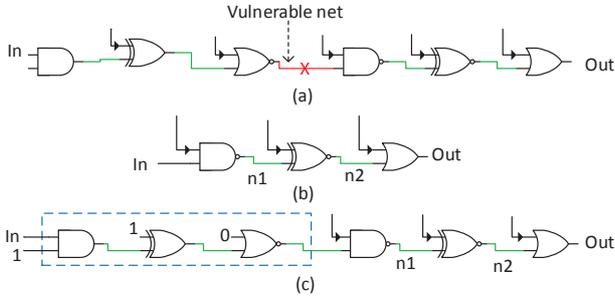}
		\caption{Generating symmetric path (a) selected suspect path  (b) a non-critical path (c) generated symmetric path to cover the vulnerable net }
		\label{fig:sym_path_gen} 
	\end{figure}

	\begin{algorithm}
		\caption{Reference path creation  -- \textit{refpathgen}()}
		\label{algo:ref_path_cretn_algo}
		\DontPrintSemicolon
		\SetAlgoLined
		\SetKwInOut{Input}{Input}
		\SetKwInOut{Output}{Output}
		\Input{Uncovered vulnerable net $n_{i}$, $S_{all\_paths}^{i}$  }
		\Output{Symmetric path pair through net $n_{i}$}
		
		$p_{short}$ = a sensitizable path with minimum delay in $S_{all\_paths}^{i}$\;
		$S_{sub}$ = a set non-critical paths $p_{sub}$ that have set of all gates in each path as subset of gates in the path $p_{short}$\ i.e. $\textit{gatesof}(p_{sub}) \subset  \textit{gatesof}(p_{short})$\;
		\If{$|S_{sub}|=0$}{
			$S_{all\_paths}^{i} = S_{all\_paths}^{i}-p_{short}$ and goto to step: 1\;
			
		}
		\ForEach{$ p_{sub}$ in $S_{sub}$ }{
			$extragates = \textit{gatesof}(p_{short}) - \textit{gatesof}(p_{sub})$\;
			Compute area overhead i.e. area of extra gates\;
		}
		Select the path $ p_{sub}$ from $S_{sub}$ with minimum area overhead\;
		$p_{ref}$ = path $p_{sub}$ with $extragates$ inserted at its input net\;
		Connect off-path inputs of $extragates$ to non-controlling values\;
		Check the functionality after extra gates insertion\;
		Check the timing of all paths through the input net of $p_{ref}$\;
		\uIf{functionality and timing are satisfied}{
			$P_{susp}^{i}= p_{short}$ and $P_{ref}^{i}= p_{ref}$\;}
		\Else{ $S_{sub}=S_{sub}-p_{sub}$ and $goto$ step: 10} 
		
		return $SymmetricPathPair_{i} \leftarrow ( P_{susp}^{i}, P_{ref}^{i} )$\;
	\end{algorithm}

	There may be some vulnerable nets for which a symmetric path pair (SPP) may not exist intrinsically in the design. We call such nets uncovered nets. To cover such nets, we create a symmetric path to be used as a reference path by adding few extra logic gates to the existing non-critical paths of the design as shown in Fig. \ref{fig:sym_path_gen}. Algorithm \ref{algo:ref_path_cretn_algo} presents the procedure for creating the reference path to cover an uncovered net $n_{i}$. We chose the shortest path through the uncovered net $p_{short}$ as the shorter paths experience less intra-die variation effects as they consist less number of gates. The non-critical paths which have gates that are a subset of gates of $p_{short}$ are collected in $S_{sub}$. The path $p_{sub}$ in $S_{sub}$ with minimum area overhead is selected and the extra gates are found as $\textit{gatesof}(p_{short}) - \textit{gatesof}(p_{sub})$. These extra gates are inserted at the input net of $p_{sub}$, thus forming a type-1 or type-2 symmetric path to be used as a reference path $p_{ref}$. The off-path inputs of extra inserted gates must be connected to non-controlling values. The functionality of the design should not change due to the addition of extra gates. Moreover, the timing of all paths passing through the input net of $p_{ref}$ must not violate any timing constraints. The designer has to ensure that all paths through the input net of the created new path must be non-critical even after the addition of extra gates and functionality must remain the same. Moreover, the created paths will be non-critical but are not redundant as they are part of the design. If the functionality and timing are satisfied $p_{short}$ and $p_{ref}$ are assigned to suspect $p_{susp}^{i}$ and reference $p_{ref}^{i}$ paths respectively. A symmetric path pair for an  uncovered net $n_{i}$ is returned as $SymmetricPathPair_{i}\leftarrow (p_{susp}^{i}, p_{ref}^{i})$.

	\section{Results}
	\label{sec:simulation_results}
	
	Simulations have been performed on ISCAS-85 benchmark circuits\cite{iscas85_benchmarks} to evaluate the proposed methodology. We carried out the experiments on a computer with an Intel Core i5-650@3.2 GHz processor and 8GB memory and the flow and proposed algorithms were implemented in python language. The benchmarks have been synthesized using Synopsys generic (SAED\_EDK32nm) library. For performing SPICE level simulations 32nm CMOS Predictive Technology Models (PTM)\cite{ptm-models} are considered. The netlists are simulated over a large number of random test vectors (i.e. $1\times10^6$) and nets with switching activity less than a predefined threshold ($1\times10^{-3}$) are considered as vulnerable nets as explained in Section \ref{subsec:potential_HT_loc}. Static timing analysis (STA) is performed using Synopsys's PrimeTime to generate path data. We used SAT-based techniques to isolate sensitizable paths out of all possible paths in the design and the SAT solver used was MiniSat-2.2. Symmetric paths are created to cover the uncovered nets as explained in Section \ref{subsec:Ref_path_creation}. Physical layout synthesis of design is carried out using Synopsys IC Compiler for generating the layout. This layout has been partitioned into $16\times16$ grids to model spatially correlated intra-die variations. We have assumed that all transistors in a gate and all the gates in a grid experience the same within-die variations. Symmetric path pairs are selected using path selection procedure explained in Section \ref{sec:path_selection_procedure}. Selected paths are extracted in SPICE netlist format from the layout of the design. Process variation on threshold voltage ($V_{th}$) is modeled as explained in Section \ref{subsec:PV}. The $3\sigma$ values for inter-die and intra-die variation are considered as $20\%$ and $15\%$ respectively \cite{Kuhn2010-32nm-pv1,Bonfiglio2013-32nm-pv2,Damrongplasit2014-32nm-pv4} . As the layout has been partitioned into $16\times16$ grids, $256$ correlated multivariate random variables are generated to model the spatial variation.  The spatial correlation is considered as 0.8 to 0.3 according to the distance between the grids. Spatial correlation decreases with an increase in distance between grids. 500 IC instances without HT and 500 IC instances with Trojan shown in Fig. \ref{fig:ht_impact_on_delay} are generated using the $V_{th}$ PV model (\ref{eq:vth_var}) and (\ref{eq:vth_intra}). HSPICE is used to perform SPICE level simulations using the generated $V_{th}$ profiles.


	\begin{table}
		\centering
		\caption{Run time of the proposed path selection algorithm for ISCAS-85 benchmark circuits (in minutes)}
		\label{table:Run-time}
		\begin{tabular}{|c|c|}
			\hline 
			Circuit&Run time\\
			\hline \hline
			c432&68 \\
			\hline
			c499&12 \\
			\hline
			c880&13 \\
			\hline
			c1355&27 \\
			\hline
			c1908&6 \\
			\hline
			c2670&14 \\
			\hline
			c3540 &765 \\
			\hline
			c5315 &143 \\
			\hline
			c7552 &285\\ 
			\hline
		\end{tabular}
	\end{table}

	
	\begin{figure*}
		\centering
		\subfloat[]{\includegraphics[scale=0.3]{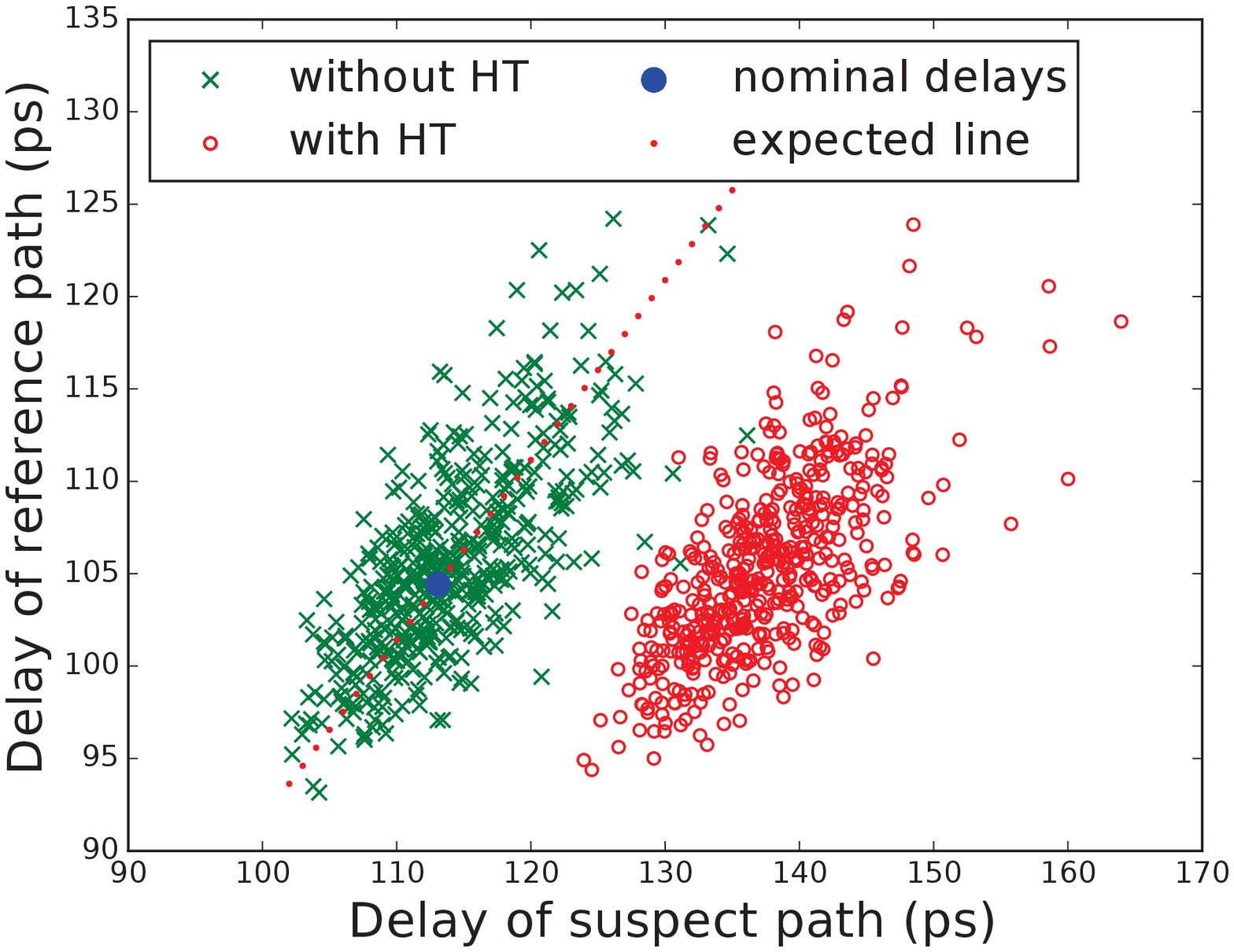}\label{fig:c432-pv}}
		\hfil
		\subfloat[]{\includegraphics[scale=0.3]{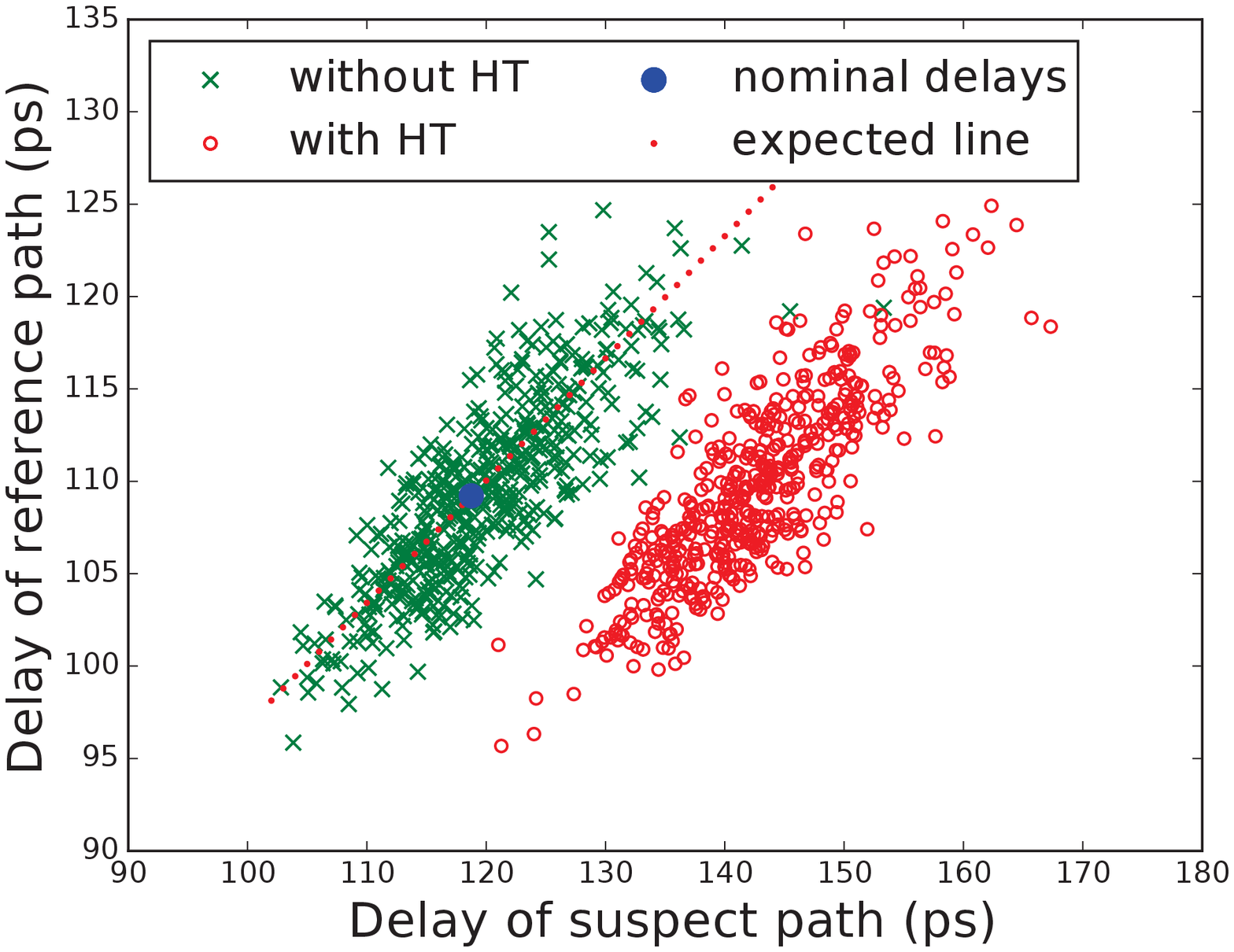}\label{fig:c1908-pv}}
		\hfil
		\subfloat[]{\includegraphics[scale=0.3]{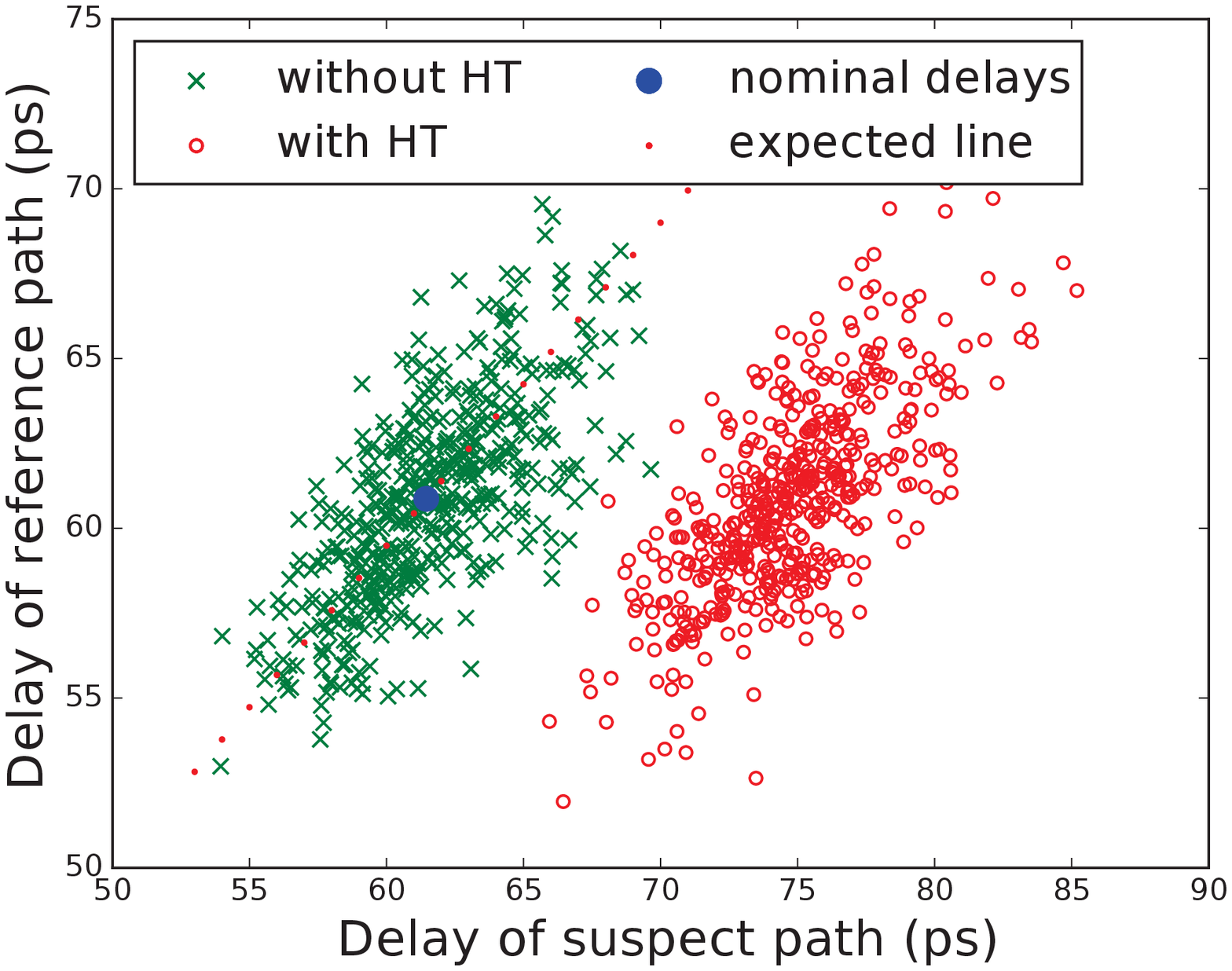}\label{fig:c5315-pv}}
		\hfil
		\subfloat[]{\includegraphics[scale=0.3]{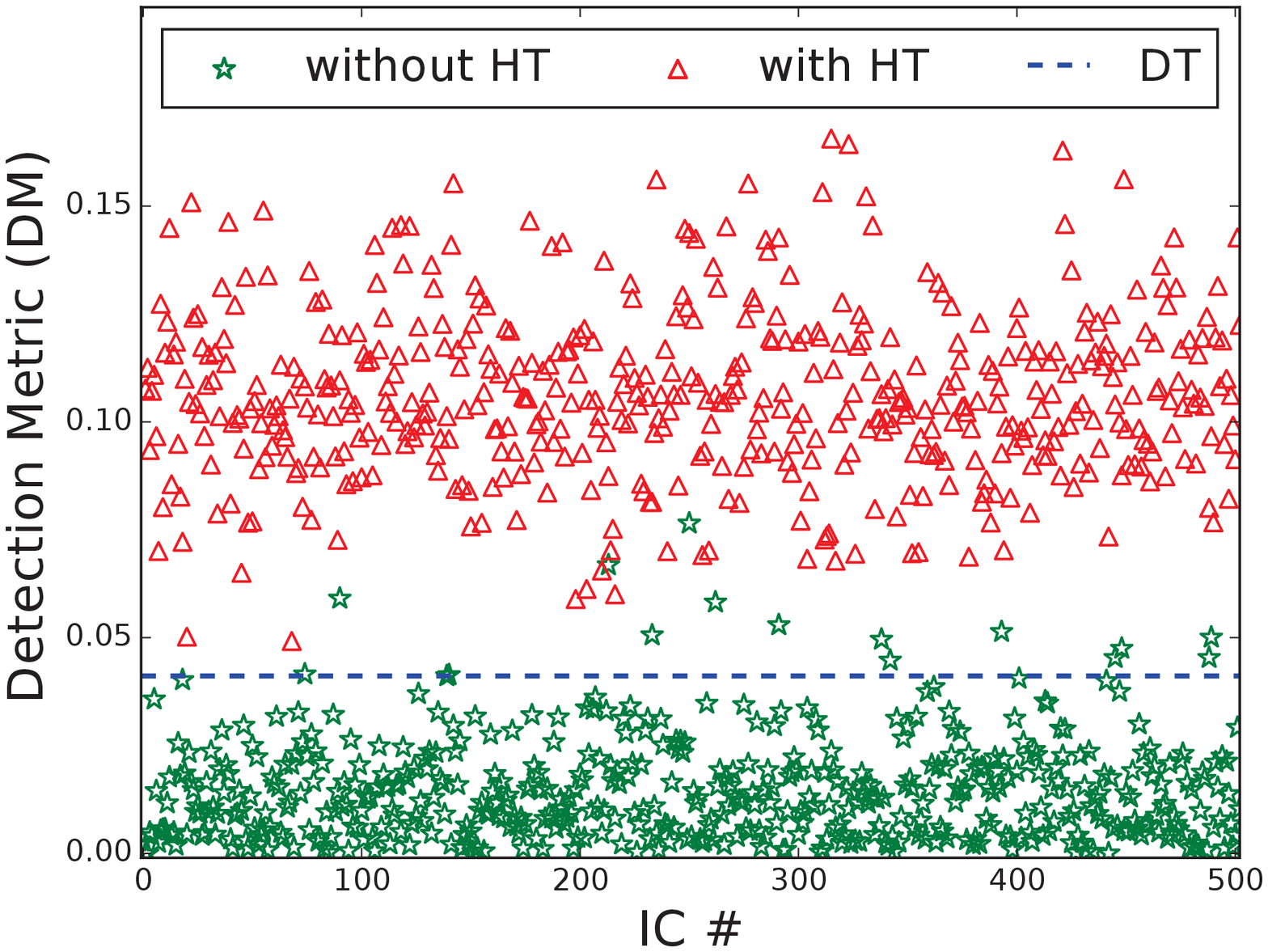}\label{fig:c432-dm}}
		\hfil
		\subfloat[]{\includegraphics[scale=0.3]{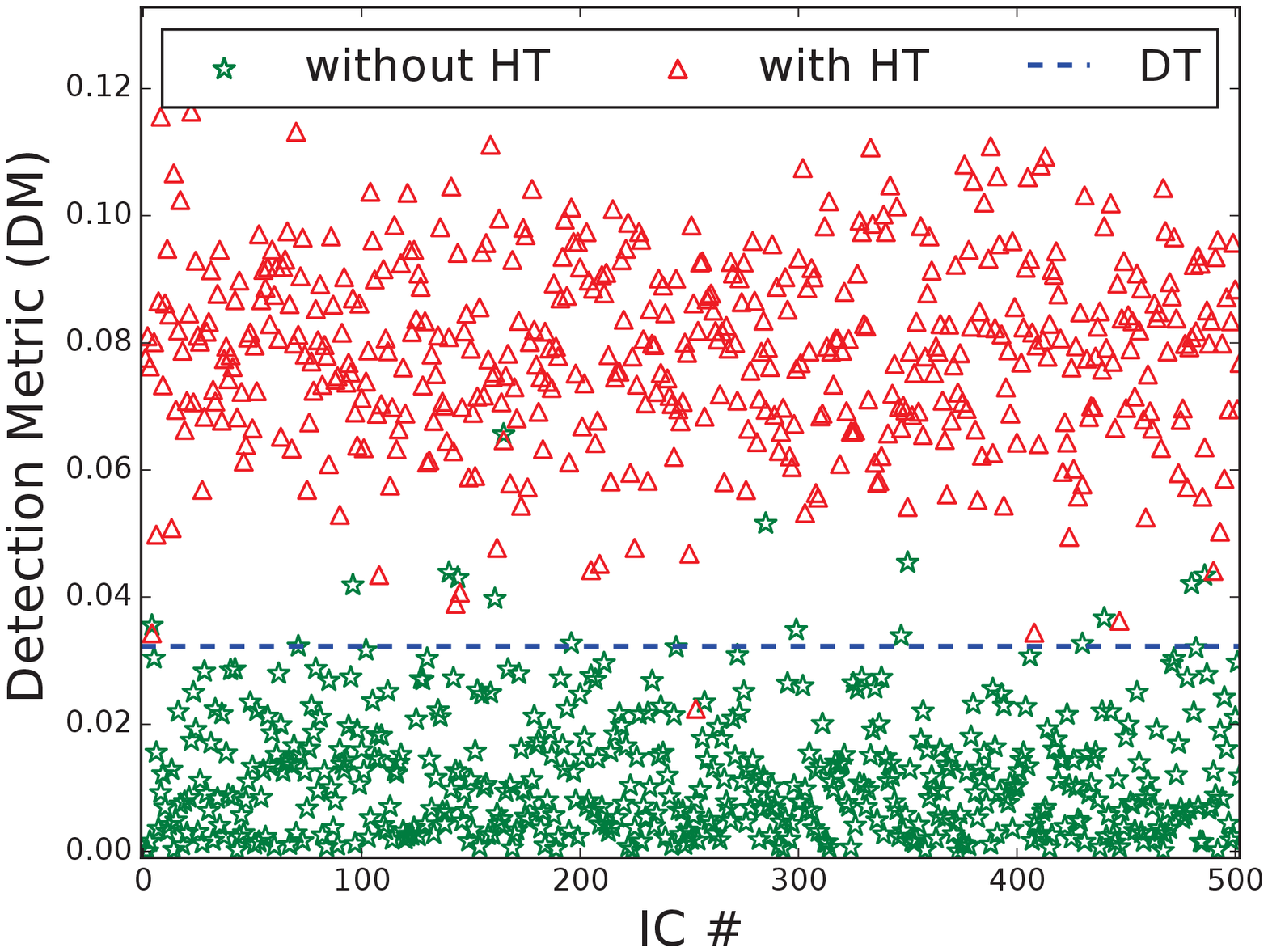}\label{fig:c1908-dm}}
		\hfil
		\subfloat[]{\includegraphics[scale=0.3]{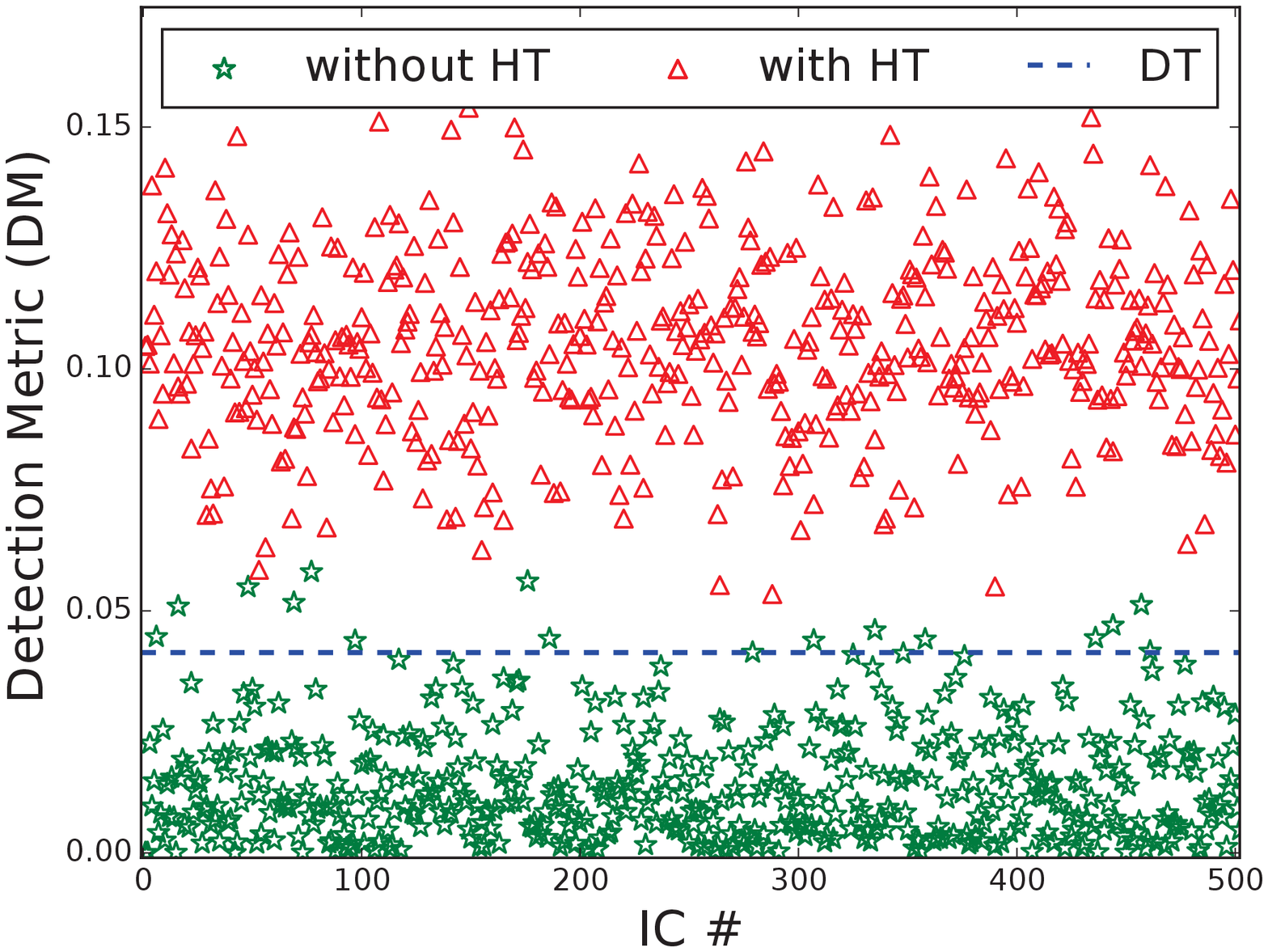}\label{fig:c5315-dm}}
		\caption{Effect of process variation and hardware Trojan on symmetric path pair of (a) c432 (b) c1908 (c) c5315 and corresponding detection metric of (d) c432 (e) c1908 (f) c5315.}
		\label{fig:benchmarks_results}
	\end{figure*}

	\begin{table*}
		\renewcommand{\arraystretch}{1.2}
		\caption{Simulation results on ISCAS-85 benchmarks}
		\label{table:sim_results}
		\centering
		\begin{tabular}{|l|c|c|c|c|c|c|c|c|c|}
			\hline
			\bfseries Circuit&\bfseries c432&\bfseries c499&\bfseries c880&\bfseries c1355&\bfseries c1908&\bfseries c2670&\bfseries c3540&\bfseries c5315&\bfseries c7552\\
			\hline\hline
			Total gates&84&159&179&188&119&240&377&556&659\\
			\hline
			Total nets&120&200&238&229&398&152&427&734&866\\
			\hline
			Sensitizable paths (\%)&44&64&98&71&94&99&49&83&77\\
			\hline
			Vulnerable nets&9&42&17&40&16&27&76&8&9\\
			\hline
			Uncovered nets&2&0&3&0&6&9&7&3&4\\
			\hline
			Extra gates added&4&0&11&0&22&14&15&9&5\\   
			\hline
			Symmetric path pairs&13&32&32&32&27&25&81&10&12\\
			\hline
			Type-1 pairs&2&30&6&32&5&12&46&3&2    \\
			\hline
			Type-2 pairs&11&2&26&0&22&13&35&7&10    \\
			\hline
			Net coverage (\%)&49.1&48.5&32.7&50.6&11.3&48.6& 55.5&3.8 &7.6\\
			\hline
			Area overhead (\%)&3.6&0&4.7&0&14.2&4.5&3.1&1.2&0.6\\
			\hline
			TPR (\%)&100&100&100&100&99&100&100&100&100\\
			\hline
			FPR (\%)&3&0&0&0&3&0&0&3&0\\
			\hline
		\end{tabular}
	\end{table*}

	The summary of our experimental results is presented in Table \ref{table:sim_results}. Rows 2 and 3 present the number of gates and nets present in each benchmark circuit considered. The percentage of sensitizable paths out of total possible paths in the design netlist is shown in row 4. We have considered only sensitizable paths for selecting symmetric path pairs using the proposed path selection algorithm. Total vulnerable nets i.e. the nets with switching activity less than the predefined threshold are presented in row 5. The uncovered nets i.e. the vulnerable nets which do not have symmetric paths intrinsically in the design are presented in row 6. We have created a symmetric path pair for each uncovered net by inserting few additional gates. The number of extra gates added to the design to create symmetric path pairs for all uncovered nets are presented in row 7. The number of symmetric path pairs (SPPs) selected to cover all the vulnerable nets is presented in row 8. We have observed that few suspect paths pass through more than one vulnerable net. Thus, the number of selected SPPs is lower than the number of vulnerable nets in c499, c1355, c2670. As we selected an SPP for every fanout branch of each vulnerable net, the number of selected SPPs is higher than the number of vulnerable nets in c432, c880, c1908, c3540, c5315, c7552. Rows 9 and 10 represent the type-1 and type-2 symmetric path pairs out of the total selected path pairs. Row 11 represents the ratio of nets covered by the selected SPPs. The net coverage ratio of c1355 is 50.6\% with no area overhead, whereas it is 11.3\% even with 14.2\% area overhead for c1908. We have observed that the net coverage ratio is higher if the majority of the paths in selected SPPs are mutually disjoint i.e. they do not share many common segments of other paths, otherwise the net coverage ratio is lower. The area overhead, due to extra gates added to the design for creating SPPs to cover uncovered nets, is presented in row 12. Finally, the true positive rate (TPR) and false-positive rate (FPR) are shown in rows 13 and 14 respectively.

	The runtime of the proposed path selection procedure for different benchmark circuits is presented in Table \ref{table:Run-time}. Our automated flow considers a set of vulnerable nets and design netlist as inputs and reports the set of selected symmetric path pairs as output. Circuit c3540 has the longer run time of 765 minutes and c1908 has the shorter run time of 6 minutes. We observe that both of these circuits have a moderate number of gates and nets in comparison to the remaining circuits (i.e. neither c3540 has the highest number of gates and nets nor c1908 has the lowest). Therefore, it can be inferred that the runtime is not dependent on the number of gates and nets present in the design. We observed that it is dependent on the total number of possible paths and the number of sensitizable paths out of the total number of possible paths in the design.

	\subsection{HT detection analysis}
	\label{subsec:HTD_analysis}
	
	The variation in delays of suspect and reference paths of a symmetric path pair (SPP) of several benchmark circuits are shown in Fig. \ref{fig:benchmarks_results}. The path delays vary along the expected straight line passing through the point of nominal delays despite inter-die variations, However, due to intra-die variations they tend to deviate away from the straight line. It is observed that the points representing the delays of SPPs in HT inserted ICs are away from the expected straight line. The graphs illustrating the effect of variation and HT on a selected SPP of some benchmarks are presented in Fig. \ref{fig:benchmarks_results}\subref{fig:c432-pv}, \ref{fig:benchmarks_results}\subref{fig:c1908-pv} and \ref{fig:benchmarks_results}\subref{fig:c5315-pv}. Detection metric ($DM$) for SPPs of each IC under test is calculated as per  (\ref{eq:DM}). The $DM$ of trojan free and with trojan ICs are shown in Fig. \ref{fig:benchmarks_results}\subref{fig:c432-dm}, \ref{fig:benchmarks_results}\subref{fig:c1908-dm} and \ref{fig:benchmarks_results}\subref{fig:c5315-dm}. The detection threshold $DT$ has been decided in a pessimistic way by allowing a false positive rate of $3\%$. The dashed horizontal lines in Fig. \ref{fig:benchmarks_results}\subref{fig:c432-dm}, \ref{fig:benchmarks_results}\subref{fig:c1908-dm} and \ref{fig:benchmarks_results}\subref{fig:c5315-dm} represent the selected $DT$ for the considered SPP of each benchmark. It can be observed that the $DM$ of few HT free ICs, is above the threshold line in Figs. \ref{fig:benchmarks_results}\subref{fig:c432-dm}, \ref{fig:benchmarks_results}\subref{fig:c1908-dm} and \ref{fig:benchmarks_results}\subref{fig:c5315-dm}. Nevertheless, the $DM$ of few Trojan inserted ICs is below the threshold line only in Fig. \ref{fig:benchmarks_results}\subref{fig:c1908-dm}. True positive rate (TPR) is defined as the ratio of number of ICs  that are detected as HT inserted out of the total number of ICs with HT inserted. False positive rate (FPR) is defined as the ratio of number of ICs that are detected as HT inserted but which in fact are HT free out of total number of ICs without HT.
	
	The TPR and FPR shown in Table \ref{table:sim_results} are for the worst-case symmetric path pair i.e the pair with lower TPR and/or higher FPR out of all selected SPPs for a benchmark under consideration. The TPR and FPR for c432 and c5315 benchmarks are 100\% and 3\%. Though the proposed method correctly detects all the Trojan inserted ICs, it falsely classifies a few Trojan free ICs as Trojan inserted ones. It is because the paths in these SPPs are not close enough to each other. The TPR and FPR of c1908 are 99\% and 3\%. Even with an FPR of 3\%, the proposed methodology cannot detect all the trojan inserted ICs and few Trojan affected ICs are identified as Trojan free. The SPPs of c432 and c1908 are intrinsic to the design, whereas the SPP of the c5315 circuit is not intrinsic but it is a created SPP to cover an uncovered net. Even though the SPP contains the shortest suspect path the created reference path is not closer to the suspect path. This leads to higher FPR in c5315. One way to improve the detection accuracy is to place the suspect and reference paths closer to each other manually while preparing the layout. However, this would incur extra area and routing overhead and requires additional design efforts.

	\subsection{Overhead analysis}
	\label{subsec:overhead_analysis}
	
	The area overhead of the proposed methodology is due to the additional gates used to create symmetric path pairs for covering uncovered nets. The benchmark circuit c1908 has the highest area overhead of 14.2\%. Whereas, the area overhead of c499 and c1355 circuits is 0\% as all the vulnerable nets are covered by the symmetric path pairs that exist in the design intrinsically. It is observed that the area overhead is dependent on the number of extra gates required to cover the uncovered nets. For circuits c1908 and c2670, even though the number of uncovered nets in c1908 is less than that of c2670, the number of extra gates required for c1908 is higher than that of c2670. Therefore, we conclude that the area overhead of the proposed method is not only dependent on the number of uncovered nets but also depends on the circuit structure.
	
	%
	\subsection{Security analysis}
	\label{subsec:security_analysis}
	
	As the proposed hardware Trojan detection (HT) method depends on the difference in path delays of two symmetric paths, it may be argued that this method may fail to detect hardware Trojans if the same Trojan gate is inserted in both the symmetric paths considered. To be able to do so, an attacker must be aware of the detection method employed which is highly unlikely. Even if the attacker knows the method, he/she has to double the number of alterations to the design to bypass the proposed method. Even if the attacker possesses the knowledge that a delay-based scheme is being used to detect HT, it is highly unlikely that he can determine the exact paths under consideration. Because there would be many paths available to pick a suspect path and for each of such selected suspect paths there would be many possible reference paths. The probability of identifying the selected path pairs for ISACS-85 circuits is shown in table \ref{table:probability}. 
	
	\begin{table}
	\centering
	\caption{Probability of an attacker identifying the selected path pair in order to bypass the proposed method}
	\label{table:probability}
	\begin{tabular}{|c|c|}
	\hline
	Circuit&Probability\\
	\hline
	c432& $3.2895\times10^{-7}$\\
	\hline
	c499& $1.0377\times10^{-6}$ \\
	\hline
	c880& $4.6964\times10^{-6}$ \\
	\hline
	c1355& $1.2752\times10^{-7}$\\
	\hline
	c1908& $5.3892\times10^{-6}$ \\
	\hline
	c2670& $2.7472\times10^{-6}$  \\
	\hline
	c3540 & $1.5922\times10^{-7}$\\
	\hline
	c5315 & $1.8503\times10^{-7}$\\
	\hline
	c7552 & $1.0382\times10^{-6}$\\ 
	\hline
	\end{tabular}
	\end{table}

	The procedure of selecting a reference path arises only during detecting Trojan. To make the advantage of self-referencing, more clear,  we compute the probability of selection of a reference path corresponding to the suspect path by the attacker while hiding the Trojan.
	
	Consider a vulnerable net of the design. Assume there are $m$ paths $P_{s1}$, $P_{s2}$,...,$P_{sm}$ passing through this net. Suppose each one  of these $m$ paths have corresponding symmetric paths from which a reference path is chosen i.e path $P_{si}$ has $k_i$ symmetric paths. The probability of the attacker finding the correct symmetric path pair is P\{ selecting the correct pair\} = $\frac{1}{\sum\limits_{i=1}^{m} k_i }$. If there are $N$ nets in the design then P\{ Attacker bypassing the method\} = $\frac{1}{N\sum\limits_{i=1}^{m} k_i }$.
	

	\section{Conclusion}
	\label{sec:conclusion}
	
	In the present work, we have considered a threat model of hardware Trojan insertion during fabrication at untrusted foundries. We have presented a novel methodology for hardware Trojan (HT) detection by analyzing the delays of topologically symmetric paths. Further, we have proposed a procedure to select paths that mitigates the effect of inter-die variation and minimizes the intra-die variation effects. We have selected topologically symmetric paths to mitigate the effects of inter-die variation and selected physically closer paths to exploit the spatial correlation for reducing the impact of intra-die variation on hardware Trojan detection accuracy. Moreover, this method uses the concept of self-referencing,  which eliminates the requirement of golden reference ICs. We have proposed and validated a procedure to create symmetric paths to cover all vulnerable nets in the design. Simulation results establish that the proposed method is able to detect HTs, which are as small as one gate. In this work, we have used simulation results to decide the detection threshold. Further work could use on-chip process variation monitors to estimate the detection threshold of each suspect IC. We are currently working on locating the trojan.

	\begin{acknowledgements} 
		This work is partly supported by the SMDP-C2SD project, sponsored by the Ministry of Electronics \& Information of Technology, Government of INDIA.
	\end{acknowledgements}

	%
	%
	\nocite{Su2020,Shang2020,Dong2020,Malik2016}
	
	\bibliographystyle{spmpsci}      
	\balance
	\bibliography{library-overleaf}    
	
	\balance

	\newpage

\end{document}